\documentclass[usenatbib]{mn2e}

\usepackage{epsfig}
\usepackage{mathrsfs,aas_macros}

\def\Lya{Ly$\alpha$~}

\def\teff{$\tau_{\rm eff}$~}
\def\HI{\hbox{H~$\rm \scriptstyle I\ $}}
\def\HII{\hbox{H~$\rm \scriptstyle II\ $}} 
\def\HeI{\hbox{He~$\rm \scriptstyle I\ $}}
\def\HeII{\hbox{He~$\rm \scriptstyle II\ $}}

\title[Evolution of the \Lya optical depth]{The evolution of the \Lya
  forest effective optical depth following \HeII reionisation}

\author[J.S. Bolton et al.] {James S.
  Bolton$^{1}$,  S. Peng Oh$^{2}$ \& Steven R. Furlanetto$^{3}$\\
  $^1$ Max Planck Institut f{\"u}r Astrophysik, Karl-Schwarzschild
  Str. 1, 85748 Garching, Germany \\
  $^2$  Department of Physics, University of California, Santa Barbara,
  CA 93106, USA\\
  $^3$Department of Physics and Astronomy, University of California,
  Los Angeles, CA 90095, USA\\}

\begin{document}

\date{17 April 2009}

\maketitle

\label{firstpage}

\begin{abstract}

Three independent observational studies have now detected a narrow
($\Delta z \simeq 0.5$) dip centred at $z=3.2$ in the otherwise smooth
redshift evolution of the \Lya forest effective optical depth.  This
feature has previously been interpreted as an indirect signature of
rapid photo-heating in the IGM during the epoch of \HeII reionisation.
We examine this interpretation using a semi-analytic model of
inhomogeneous \HeII reionisation and high resolution hydrodynamical
simulations of the \Lya forest.  We instead find that a rapid ($\Delta
z \simeq 0.2$) boost to the IGM temperature ($\Delta T \simeq
10^{4}\rm~K$) beginning at $z=3.4$ produces a well understood and
generic evolution in the \Lya effective optical depth, where a sudden
reduction in the opacity is followed by a gradual, monotonic recovery
driven largely by adiabatic cooling in the low density IGM.  This
behaviour is inconsistent with the narrow feature in the observational
data.  If photo-heating during \HeII reionisation is instead extended
over several redshift units, as recent theoretical studies suggest,
then the \Lya opacity will evolve smoothly with redshift.   We
conclude that the sharp dip observed in the \Lya forest effective
optical depth is instead most likely due to a narrow peak in the
hydrogen photo-ionisation rate around $z=3.2$, and suggest that it may
arise from the modulation of either reprocessed radiation during \HeII
reionisation, or the opacity of Lyman limit systems.  

\end{abstract}
 


\section{Introduction}

Observations of the \HeII \cite{GunnPeterson65} trough in the spectra of
intermediate redshift quasars, coupled with large fluctuations in the
mean transmission of the \HeII \Lya forest, provide the most direct
evidence for the tail-end of \HeII reionisation occurring around $z=3$
(\citealt{Jakobsen94,Davidsen96,Heap00,Zheng04b,Shull04,Fechner06}). However,
to date only a handful of ultraviolet (UV) quasar spectra suitable for
detailed analyses of the \HeII \Lya forest have been obtained
(although see \citealt{Syphers08}).  Moreover, the \Lya transmission
is sensitive to small \HeII fractions only, and is thus unsuitable for
probing the earliest stages of \HeII reionisation.

Fortunately, the impact of \HeII reionisation on the intergalactic
medium (IGM) can still be probed, albeit indirectly, over a wider
redshift baseline with the existing wealth of high quality, high
signal-to-noise \HI \Lya forest data.  Firstly, the increased
temperature of the IGM expected following \HeII reionisation
(\citealt{MiraldaRees94,AbelHaehnelt99,Paschos07}) will thermally broaden
absorption lines in the \Lya forest (\citealt{Haehnelt98}).  There
is some evidence that observed line widths are consistent with a
sudden increase in the IGM temperature around $z=3.3$
(\citealt{Schaye00,Ricotti00}), although the error bars on the
measurements are large and not all studies agree on this result
(\citealt{McDonald01}).  Secondly, the residual \HI fraction in the
IGM, which is in photo-ionisation equilibrium with the metagalactic UV
background, will be lowered through the temperature dependence of the
\HII recombination coefficient ($n_{\rm HI} \propto T^{-0.7}$).  If
the temperature of the IGM rises suddenly following \HeII
reionisation, a similarly sudden decrease in the \HI fraction, and
hence the observed \Lya forest opacity, will result
(\citealt{Theuns02d,Bernardi03,Faucher08}).

\cite{Bernardi03}, in a study using 1061 moderate resolution quasar
spectra obtained from the Sloan Digital Sky Survey (SDSS),
statistically measured such a departure from the otherwise smooth,
power law evolution of the \Lya forest opacity.  A sudden decrease of
around $10$ per cent was observed in the effective optical depth at
$z=3.2$, followed by a recovery to its former power law evolution by
$z=2.9$.  \cite{Faucher08} have subsequently reconfirmed the
\cite{Bernardi03} result by using 86 high resolution, high
signal-to-noise spectra obtained with Keck/HIRES/ESI and Magellan/MIKE
to directly measure the \Lya opacity.  Furthermore, \cite{Dallaglio08}
have also recently detected a narrow dip in the \Lya forest effective
optical depth at $z=3.2$, albeit at a low level of statistical
significance ($2.6\sigma$), using another set of 40 high resolution
spectra obtained with VLT/UVES. 

\cite{Theuns02d} (hereafter T02) compared detailed hydrodynamical
simulations of the \Lya forest to the \cite{Bernardi03} data and
interpreted this feature as indirect evidence for photo-heating during
\HeII reionisation.  However, recent theoretical studies indicate it is
difficult to explain this narrow feature in the \Lya opacity evolution
by invoking a rapid temperature boost in the IGM alone.
\cite{Bolton09} used analytical and numerical arguments to
demonstrate that there are too few hard photons available to heat the
entire IGM by the required amount ($\sim 10^{4}\rm~K$) over the
timescale on which the observed opacity decreases ($\Delta z =
0.1-0.2$).  State-of-the-art radiative transfer simulations also
indicate that \HeII reionisation and reheating will be an extended
rather than sudden process, with the volume averaged IGM temperature
rising gradually from higher redshifts (\citealt{McQuinn09}).  A
rapid injection of energy into the IGM during \HeII reionisation is
thus unlikely to explain the sharp initiation of the observed dip in
the \Lya forest opacity. 

Somewhat separate to this argument, however, is the explanation for
the subsequent {\it recovery} of the observed \Lya opacity to its
former power law evolution by $z=2.9$.  Let us assume that the narrow
dip in the \Lya forest effective optical depth is indeed initiated by
a large, rapid temperature boost in the IGM following \HeII
reionisation (but see \citealt{Bolton09,McQuinn09}).  Although this
will not occur over the whole IGM, such rapid heating is not
necessarily excluded in localised patches close to quasars with very
hard spectra, where the requisite hard photons are abundant.
Following reionisation, the thermal evolution of the IGM at low density
is dominated by the balance between photo-heating and adiabatic
cooling due to Hubble expansion (\citealt{HuiGnedin97,Theuns98}).
However, if the \Lya opacity evolution were driven by the thermal
state of the IGM alone through the \HII recombination rate, any
subsequent recovery in the opacity due to adiabatic cooling would
occur over a Hubble time -- too long to explain the rapid recovery in
the \cite{Bernardi03} feature.  The study of T02 instead demonstrated
that this interpretation is too simplistic.  They found that the
hydrodynamical response of the IGM following a sudden reheating,
coupled with the impact of the extra electrons liberated during \HeII
reionisation, are enough to drive the \Lya opacity back to its former
power law evolution by $z=2.9$ as observed by \cite{Bernardi03} --
much earlier than one would expect using the simple argument above. 

To date, the only dedicated study of the impact of \HeII reionisation
on the \Lya forest effective optical depth using high resolution
hydrodynamical simulations of the IGM has been performed by T02.
\cite{McQuinn09} recently presented a detailed analysis of the
evolution of the \Lya forest opacity using radiative transfer
simulations of \HeII reionisation.  However, their study necessarily
used low resolution, post-processed dark matter simulations, and they
were thus unable to model the impact of the gas hydrodynamics on the
recovery of the \Lya forest opacity.  In light of these recent
observational and theoretical results we therefore re-examine the
impact of \HeII reionisation on the evolution of the \Lya forest
opacity using semi-analytic modelling and high resolution
hydrodynamical simulations of the IGM.  In particular, we shall focus
on the explanation for the {\it recovery} of the dip observed in the
\Lya forest opacity.  This paper is therefore closely related to the
work presented in \cite{Bolton09}, where the issues surrounding the
{\it initiation} of the feature first detected by \cite{Bernardi03}
({\it i.e.}  the plausibility, or lack of, for the rapid photo-heating
of the IGM) were examined in detail. 

The structure of this paper is as follows.  We begin in \S2 with a
brief review of the relationship between the \Lya forest opacity and
the underlying physical properties of the IGM.  In \S3 we use a
semi-analytic model to examine the evolution of the \Lya forest
opacity following photo-heating in the IGM during inhomogeneous \HeII
reionisation.  Motivated by these results, we then proceed to model
the \Lya forest opacity in more detail using hydrodynamical
simulations of the IGM.  The simulations are described in \S4, and the
evolution of the \Lya forest opacity in the simulations is presented
in \S5.    In \S6 we investigate the impact of sudden reheating on the
IGM gas distribution and peculiar velocity field; we find that our
simulations are unable to reproduce the narrow feature observed in the
\Lya forest opacity evolution.  Finally, we consider alternative
explanations for the observational data in \S7 before concluding in
\S8.


\section{The fluctuating Gunn-Peterson approximation}

The forest of \Lya absorption lines observed in the spectra of high
redshift quasars originates from the neutral hydrogen remaining in the
intervening, low density IGM following \HI reionisation
(\citealt{Bi92,Zhang95,Hernquist96,MiraldaEscude96,Theuns98}).  The
mean normalised flux of the \Lya forest, $\langle F \rangle=
\langle I_{\rm observed}/I_{\rm emitted}\rangle$, is the simplest
observable quantity, and it is often expressed as an effective
optical depth

\begin{equation} \tau_{\rm eff} = -\ln \langle F \rangle \equiv - \ln
  \langle e^{-\tau} \rangle, \end{equation}

\noindent
where $\tau$ is the underlying \Lya optical depth in each pixel of the
spectrum or spectra from which $\langle F \rangle$ is measured.
Assuming the IGM is highly ionised and in photo-ionisation equilibrium
with the metagalactic UV background, and the low density IGM
($\Delta=\rho/\langle \rho \rangle \leq 10$) follows a power-law
temperature density relation, $T=T_{0}\Delta^{\gamma-1}$
(\citealt{HuiGnedin97,Valageas02}), the \Lya optical depth at $z\ga 2$
may be written as ({\it e.g.}
\citealt{Weinberg99,McDonaldMiraldaEscude01})

\[ \tau \simeq 1.0 \frac{(1+\chi_{\rm
      He})}{\Gamma_{-12}}
  \left(\frac{T_{0}}{10^{4}\rm~K}\right)^{-0.7}\left(\frac{\Omega_{\rm
      b}h^{2}}{0.024}\right)^{2}\left(\frac{\Omega_{\rm
      m}h^{2}}{0.135}\right)^{-1/2} \]
\begin{equation} \hspace{7mm} \times
  \left(\frac{1+z}{4}\right)^{9/2}\Delta^{2-0.7(\gamma-1)}, \label{eq:FGPA} \end{equation}

\noindent
where $\Omega_{\rm b}$ and $\Omega_{\rm m}$ are the present day baryon
and matter densities as a fraction of the critical density,
$h=H_{0}/100\rm~km~s^{-1}~Mpc^{-1}$ for the present day Hubble
constant $H_{0}$, $\Delta=\rho/\langle \rho \rangle$ is the normalised
gas density, $T_{0}$ is the gas temperature at mean density, $\gamma$
is the slope of the temperature density relation and
$\Gamma_{-12}=\Gamma_{\rm HI}/10^{-12}\rm~s^{-1}$ is the hydrogen
photo-ionisation rate.  The power-law temperature dependence is due to
the case-A \HII recombination coefficient, and  $\chi_{\rm He}$
accounts for the extra electrons liberated during \HeII reionisation;
$\chi_{\rm He}=1.08$ prior to \HeII reionisation and $\chi_{\rm
  He}=1.16$ afterwards for a helium fraction by mass of $Y=0.24$
(\citealt{OliveSkillman04}).  The effective optical depth can then be
estimated by integrating over all possible IGM densities $\Delta$.

Eq.~(\ref{eq:FGPA}) is the Fluctuating Gunn-Peterson Approximation
(FGPA), and although it ignores the effect of redshift space
distortions on the \Lya forest opacity, it clearly elucidates the
relationship between the opacity and the underlying physical
properties of the IGM.  A sudden decrease in the \Lya effective
optical depth ({\it e.g.}  \citealt{Bernardi03}) can thus be
attributed to an increase in the IGM temperature, either by raising
$T_{0}$ or changing\footnote{Lowering (raising) the value of $\gamma$
  will increase the temperature in underdense (overdense) regions of
  the IGM while decreasing the temperature in overdense (underdense)
  regions.  Whether raising or lowering $\gamma$ subsequently
  decreases the effective optical depth of the \Lya forest thus
  depends on redshift.  At $z \geq 3$ the \Lya forest opacity is
  mainly sensitive to underdense regions in the IGM.  Lowering
  $\gamma$ therefore decreases the \Lya effective optical depth at $z
  \geq 3$ by producing hotter, more highly ionised voids ({\it e.g.}
  \citealt{Bolton05,Faucher08b}). Typical values adopted for $\gamma$
  in \Lya forest models lie in the range $1 \leq \gamma \leq 1.6$,
  although there is some evidence to suggest a more complex
  relationship between temperature and density, perhaps with
  $\gamma<1$,  may be required to reproduce the observed \Lya forest
  flux distribution (\citealt{Becker07,Bolton08}).}  
$\gamma$, an increase in the photo-ionisation rate, or a combination
of both.  On the other hand, an increase in the free electron fraction
will raise the opacity by reducing the recombination timescale.  Both
an increase in the IGM temperature and an $8$ per cent change in the
free electron fraction will be associated with \HeII reionisation.
Keeping these points in mind, we now proceed to describe a
semi-analytical model for inhomogeneous \HeII reionisation which
utilises this useful approximation.


\section{The \Lya forest opacity during inhomogeneous \HeII reionisation}
\label{semi-analytic}

\subsection{A semi-analytic model}

Before analysing our detailed hydrodynamical simulations, we first
consider a simplified semi-analytic model which illustrates the
difficulty of reproducing the localised feature observed by
\cite{Bernardi03} in the \Lya forest.  Although this model is
unable to compute many of the detailed properties of the \Lya forest
(such as line broadening and the impact of peculiar velocities), it
has the virtue of including inhomogeneous reionisation
\citep{FurlanettoOh08b}, which our hydrodynamical simulations do not.
The model is based on the calculations presented in
\citet{FurlanettoOh08}, who examined how \HeII reionisation can
affect the temperature-density relation of the IGM, and we refer the
reader there for more details.

In brief, the model has three parts.  The first determines the
reionisation history of gas elements of a given density in the IGM
\citep{FurlanettoOh08}.  For most of our calculations, we assume that
the probability that an element is reionised at any particular time is
independent of its density, simply tracing the mean ionisation
history, $\bar{x}_i(z)$.  We also compare to a ``density-driven" model
in which high-density regions are ionised first (because they sit near
the biased regions that host quasars).  However, recent numerical
simulations of \HeII reionisation suggest that the ionised regions
are large and rare enough that this correlation is weak
\citep{McQuinn09}, so we usually use the density-independent model.
We will assume that \HeII reionisation ends at $z_{\rm He}=3.2$,
consistent with the calculations below, and that hydrogen reionisation
(which only affects the temperature of gas for which helium is still
singly-ionised) occurs at $z_{\rm H}=8$; the latter has little effect
on our calculation.

To make this comparison as straightforward as possible, we assume that
$\bar{x}_i(z)$ is simply proportional to the mass in galaxies with $m
\ga 5 \times 10^{11}\rm \, M_\odot $ (in other words, these massive
galaxies contain supermassive black holes that have gone through
bright quasar phases).  This provides a reasonable approximation to
the quasar emissivity (compared to, {\it e.g.}, the luminosity
function of \citealt{Hopkins07}) and also leads to fast \HeII
reionisation, with over $70\%$ of the ionisation occurring after
$z=4.2$, if \HeII reionisation ends at $z=3.2$
\citep{FurlanettoOh08b}.

The second ingredient is to follow the thermal evolution of each gas
element after reionisation.  We again use the method presented in
\citet{FurlanettoOh08}, which is in turn based on \citet{HuiGnedin97}.
We include all of the relevant atomic cooling, heating, recombination,
and photo-ionisation processes, as well as adiabatic expansion (and
collapse for overdense regions).  After reionisation, the dominant
processes are photo-heating and adiabatic cooling, which together
determine the asymptotic, nearly power law, temperature-density
relation \citep{HuiGnedin97,HuiHaiman03}.  

The chief uncertainty in this model is the initial temperature after
reionisation, $T_i$, which depends in a non-trivial manner on the
spectrum of the photons ionising the gas parcel
\citep{AbelHaehnelt99,Tittley07,Paschos07,Bolton09,McQuinn09}.  This
may itself be inhomogeneous in the IGM, because low-energy photons
will be absorbed near their hosts, leaving many of the ionisations to
be done by hard photons \citep{AbelHaehnelt99}.  We therefore consider
a range of possibilities for $T_i$.  Our fiducial model takes $T_i=4
\times 10^4$~K, larger than is typically expected in order to
exaggerate the effect of heating ({\it e.g.}, \citealt{Bolton09, McQuinn09}).

Finally, in order to compute the average optical depth we need to
assume an IGM density distribution, $P_V(\Delta)$ (averaged by
volume), as well as the distribution
of temperatures and ionisation rates at each density.  We will use the
density distribution recommended by \citet{MiraldaEscude00}:

\begin{equation}
P_V(\Delta) \, d \Delta = A_0 \Delta^{-\beta} \exp \left[ -
  \frac{(\Delta^{-2/3} - C_0)^2}{2(2 \delta_0/3)^2} \right] \, d
\Delta.
\label{eq:pvd}
\end{equation}

\noindent
This form fits cosmological simulations at $z=2$--$4$ quite well.
Note that the underlying simulation had somewhat different
cosmological parameters than the currently preferred values; it is,
however, accurate enough for the qualitative calculations that follow.

\subsection{The \Lya effective optical depth}

To compute the \Lya effective optical depth, we assign each volume
element of known density an optical depth according to
Eq.~(\ref{eq:FGPA})\footnote{We drop the assumption of a power law
  temperature-density relation in this instance; see
  \citet{FurlanettoOh08} for some example temperature distributions.}.
Our model makes no predictions about the amplitude of the ionising
background, and for simplicity we set $\Gamma_{-12}=1$ over the entire
redshift range $z=2$--$5$.  This is consistent with constraints
derived by comparing observational data to simulations of the \Lya
forest at $2\leq z \leq 4$ (\citealt{Bolton05}), and yields a mean
transmission reasonably close to the observed values.

We choose the temperatures by following gas elements of the
appropriate density after their most recent reionisation event using
the thermal evolution code described above, with the reionisation
redshifts distributed according to the overall ionisation history.  We then
calculate

\begin{equation}
e^{-\tau_{\rm eff}} = \int_0^\infty d\Delta \, P_V(\Delta) \int dT \,
P(T|\Delta) e^{-\tau(T,\Delta)},
\label{eq:taueff-sa}
\end{equation}

\noindent
where $P(T|\Delta)$ is the probability distribution of temperatures
for elements at density $\Delta$ and \teff is the effective optical
depth.  Again, we emphasise that this ignores peculiar velocities of
the gas, line blending (and indeed the wings of every line), and the
clustering of the absorbers, but it provides a qualitative description
of the evolving transmission.

\begin{figure}
\begin{center}

  \includegraphics[width=0.42\textwidth]{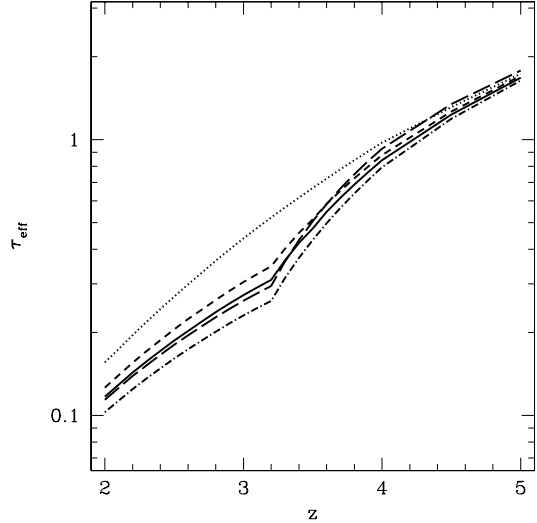}
\vspace{-0.2cm}        
\caption{Mean transmission histories in our semi-analytic models.  The
  dotted curve shows $\tau_{\rm eff}(z)$ if we ignore \HeII
  reionisation and set $T=20,000$~K throughout.  The other curves include
  \HeII reionisation at $z_{\rm He}=3.2$, as described in the text.
  The short-dashed, solid, and dot-dashed curves assume $T_i=(3,4,6) \times
  10^4$~K, respectively.  The long-dashed curve assumes $T_i=4 \times
  10^4$~K but that reionisation proceeds from high to low densities.}

\label{fig:taueff-analytic}
\end{center}
\end{figure}

Fig.~\ref{fig:taueff-analytic} shows some example histories computed
in this manner.  As a basis for comparison, the dotted curve ignores
\HeII reionisation and assumes a constant ionising background and IGM
temperature ($T=20,000$~K).  This does not quantitatively match the
observed evolution; it merely serves to show that, without \HeII
reionisation, our model produces smooth, nearly featureless evolution.  

The other curves assume that helium reionisation ends at $z_{\rm
  He}=3.2$.  The short-dashed, solid, and dot-dashed curves assume
$T_i=(3,4,6) \times 10^4$~K, respectively (all relatively large in
order to exaggerate the effect; see \citealt{Bolton09, McQuinn09} for
self-consistent estimates of the plausible temperature range).  The
long-dashed curve takes $T_i=4 \times 10^4$~K and assumes that
reionisation is density-dependent \citep{FurlanettoOh08}; this aspect
of reionisation has only a small effect on the mean transmission.
This is because the \Lya forest is sensitive to only a narrow range of
densities near the mean, where the density modulation is modest
anyway.

According to these models, \HeII reionisation can certainly induce a
feature in $\tau_{\rm eff}$, so long as the temperature increase is
large enough.   In all cases, $\tau_{\rm eff}$ falls relatively
steeply until $z_{\rm He}$ before levelling off and returning closer
to the expected evolution without \HeII reionisation.   The shape is
generic within these reionisation models, although note that the
downward turn in \teff is still significantly shallower than that
observed in the observational data (\citealt{Bernardi03,Faucher08}).  The
pre-reionisation phase can be steepened by making reionisation occur
faster, but the model assumed here is not far from empirical estimates
of the evolution of the quasar emissivity.  Moreover, compressing
\HeII reionisation into a short time interval limits the temperature
jump that it can induce, because the photo-ionisation timescale
associated with high-energy photons (which provide the most efficient
heating) is quite long \citep{Bolton09}.

However, the post-reionisation evolution is completely generic and
easy to understand.  In this regime, the thermal evolution is
dominated by the competition between photo-heating and cooling by
adiabatic expansion.  The timescale for the cooling is therefore the
expansion, or Hubble, time --  a substantial fraction of the age of
the Universe.  There is no way to avoid this behaviour for the
temperature history, and to the extent that the Ly$\alpha$ forest
depends only on these temperatures, \HeII reionisation cannot induce a
\emph{narrow} feature in the transmission which recovers quickly to
its pre-reionisation evolution.

Of course, we have emphasised that the forest is indeed more complex
than this model, because of peculiar velocities, geometric effects,
and line blending.  In principle, these can induce a narrower
feature, and T02 appealed to just such an effect to explain the
\cite{Bernardi03} feature.  In their simulations, once the reheated
gas in the IGM became overpressurised with respect to its surroundings
it started to expand, resulting in a sudden change in the peculiar
velocity gradients in the IGM.  This extended the \Lya absorption
lines in redshift space by shifting absorption from the saturated line
cores to the wings, increasing the line equivalent widths and hence
the mean \Lya opacity.   In the remainder of this paper, we will
examine all of these effects in more detail using new hydrodynamical
simulations that do include a detailed reconstruction of the
Ly$\alpha$ forest.


\section{Hydrodynamical simulations of the \Lya forest}
\subsection{Initial conditions}

The hydrodynamical simulations in this study are performed using
an upgraded version of the
publicly available parallel Tree-SPH code GADGET-2
(\citealt{Springel05}).   All simulations have a box size $15h^{-1}$
comoving Mpc and contain $2 \times 400^{3}$ gas and dark matter
particles.  The mass of each gas particle is $9.4\times
10^{5}\rm~M_{\odot}$ and the gravitational softening length is
$1/30^{\rm th}$ of the mean linear interparticle spacing.  This
adequately resolves the \Lya forest at $2 \leq z \leq 4$
(\citealt{Theuns98,Bolton08}) and provides just over twice the mass
resolution of the simulations used in the T02 study.\footnote{T02 use
  a modified version of Hydra (\citealt{Couchman95}) to run
  hydrodynamical simulations in a $12h^{-1}$ comoving Mpc box with a
  gas particle mass of $2.0\times 10^{6}\rm~M_{\odot}$.}  Star
formation is included using a simplified prescription which converts
all gas particles with overdensity $\Delta > 10^{3}$ and temperature
$T<10^{5}\rm~K$ into collisionless stars, significantly speeding up
the simulations.  Outputs are saved every $\Delta z=0.05$ in the
redshift range $2 \leq z \leq 4$, enabling a very fine sampling of the
simulation data with redshift.

\begin{figure*}
\centering 
\begin{minipage}{180mm} 
\begin{center}
\psfig{figure=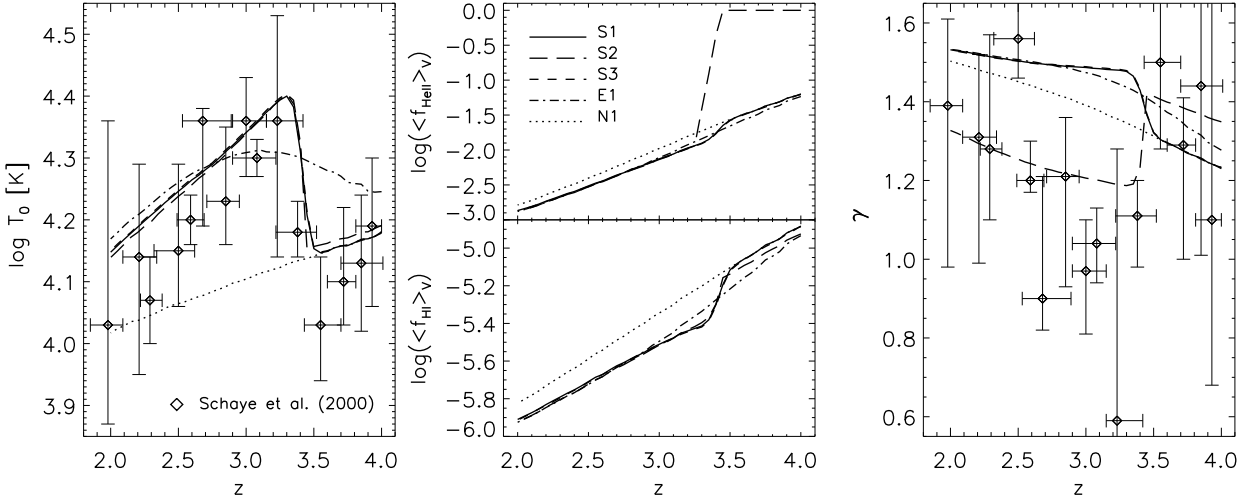,width=1.0\textwidth}
\vspace{-0.5cm}
\caption{{\it Left:} The temperature of the IGM at mean density,
  $T_{0}$, in the five simulations used in this study: S1 (solid
  curve), S2 (long-dashed curve), S3 (short-dashed curve), E1
  (dot-dashed curve) and N1 (dotted curve).  The temperatures are
  given at intervals of $\Delta z = 0.05$ over the redshift range $2
  \leq z \leq 4$.  These data are compared to observational constraints
  obtained by \citet{Schaye00} from an analysis of absorption line
  widths in the \Lya forest. {\it Centre:} The corresponding volume
  weighted \HeII (upper panel) and \HI (lower panel) fractions.  {\it
    Right:} The slope of the power law temperature-density relation,
  $T=T_{0}\Delta^{\gamma-1}$.  The values are obtained by a least
  squares power law fit to the volume weighted temperature-density
  plane in the simulations at $0.1 \leq \Delta \leq 1$ and $T \leq
  10^{5}\rm~K$. The data are again compared to observational
  constraints from \citet{Schaye00}.  Note that the S1 and S3 models
  are almost indistinguishable in all three panels.}
\label{fig:ionisation} 
\end{center} 
\end{minipage}
\end{figure*}

The ionisation state of the gas particles is computed in the optically
thin limit using a non-equilibrium ionisation algorithm which follows
the abundances of six species: H~$\rm \scriptstyle I$, H~$\rm
\scriptstyle II$, He~$\rm \scriptstyle I$, He~$\rm \scriptstyle II$,
He~$\rm \scriptstyle III$ and $e^{-}$ (\citealt{BoltonHaehnelt07}).
The ultraviolet background (UVB) model of \cite{HaardtMadau01}
(hereafter HM01) including emission from quasars and galaxies is used
to compute the H~$\rm \scriptstyle I$, \HeI and \HeII photo-ionisation
and heating rates.  This model is in good agreement with observational
constraints on the metagalactic \HI photo-ionisation rate
(\citealt{Tytler04,Bolton05}), although we set the HM01
photo-ionisation and heating rates for all species to zero at $z>6$ in
order to match the simulations of T02.

The UVB is assumed to be spatially uniform in all our simulations.
This will be a reasonable approximation at the \HI photo-ionisation
edge at $z<4$, when the mean free path for hydrogen ionising photons
is much larger than the average separation between ionising sources
(\citealt{BoltonHaehnelt07b,Faucher08b,Furlanetto08b}).  However, a
full radiative transfer implementation is required to model the impact
of inhomogeneous \HeII reionisation on the \Lya forest opacity ({\it
  e.g.}  \citealt{Maselli05,Tittley07,Paschos07,Bolton09,McQuinn09}).
Although \HeII reionisation does not {\it directly} impact \Lya forest
effective optical depth, our simulations do not capture the effect of
the resulting fluctuations in the IGM temperature, which do impact on
the effective optical depth through the temperature dependence of the
\HII recombination coefficient.  However, our simulation volume is
$15h^{-1}$ comoving Mpc, which is small on comparison to the $\sim 50$
comoving Mpc scales on which large temperature fluctuations are
expected (\citealt{McQuinn09}).  Although not ideal, our assumption of
instantaneous reheating on this smaller scale is therefore a
reasonable approximation.  Furthermore, note that HM01 implicitly
assumes that \HeII is reionised at all redshifts, {\it i.e.} \HeII
only exists in the dense \Lya absorbers which they model. Radiative
transfer effects obviously modify the form of the UV background prior
to \HeII reionisation ({\it e.g.} \citealt{MadauHaardt08}), both in
\HeII ionising radiation above 4 Ry and in reprocessed radiation from
\HeII Lyman series and two photon emission. Thus, the UV background
could be quite different before reionisation is complete. We comment
on this more in \S \ref{section:ionizing_background}. 

The simulations were all started at $z=99$, with initial conditions
generated using the transfer function of \cite{EisensteinHu99}.  The
cosmological parameters adopted are $\Omega_{\rm m}=0.26$,
$\Omega_{\Lambda}=0.74$, $\Omega_{\rm b}h^{2}=0.024$, $h=0.72$,
$\sigma_{8}=0.85$ and $n_{\rm s}=0.95$.  These are consistent with the
fifth year {\it Wilkinson Microwave Anisotropy Probe} (WMAP) data
(\citealt{Dunkley08}), aside from a slightly larger value for the
matter power spectrum normalisation.  The gas is assumed to be of
primordial composition with a helium mass fraction of $Y=0.24$ ({\it
  e.g.} \citealt{OliveSkillman04}).

\subsection{Thermal and ionisation histories}

Five hydrodynamical simulations, listed in Table~\ref{tab:sims}, were
performed for this study.  The simulations use different models for
the thermal and ionisation evolution of the IGM, but aside from the S3
model, which we shall discuss later, are identical in all other
respects.  The different thermal histories are constructed by
increasing the HM01 \HeII photo-heating rate in the simulations ({\it
  e.g.}  T02;~\citealt{Bolton05,Jena05}).  This mimics spectral
hardening due to radiative transfer effects during \HeII reionisation
by boosting the mean excess energy per \HeII photo-ionisation
(\citealt{AbelHaehnelt99,Bolton04}).  Note that we only model the jump
in heating rates, which affects the thermal evolution; apart from
model S2 which we discuss below, we do not model the jump in the \HeII
fraction itself.

The IGM temperatures at mean density, $T_{0}$, in all five simulations
are displayed in the left panel of Fig.~\ref{fig:ionisation} as a
function of redshift.  These data are compared to observational
constraints obtained by \cite{Schaye00} from an analysis of absorption
line widths in the \Lya forest.  Note that these thermal histories are
not fits to the observational data; they are instead merely chosen to
be representative of IGM thermal histories presented in the
literature.  The S1 model (solid curve) closely resembles the
$T_{0}$ evolution in the simulations of T02, with a sudden increase
($\Delta z = 0.1$) in temperature, $\Delta T\sim 10^{4}\rm~K$,
beginning at $z=3.4$.  Note that this temperature boost occurs over a
much shorter timescale than that in our semi-analytical model of
\HeII reionisation (\citealt{FurlanettoOh08}).  The E1 model
(dot-dashed curve) is qualitatively similar to recent results from
detailed three dimensional radiative transfer simulations of \HeII
reionisation (\citealt{McQuinn09}).  The temperature boost develops
over a longer timescale as quasars gradually photo-heat the IGM.  
The third simulation, N1 (dotted curve), is included as a control
model, and has no temperature boost.

\begin{table}
\centering
\caption{Hydrodynamical simulations used in this study.  All the
  simulations have a box size of $15h^{-1}$ comoving Mpc and contain
  $2\times 400^{3}$ gas and dark matter particles.}
\begin{tabular}
{c|c}
  \hline
    Model & Thermal history description  \\  
  \hline
 S1            & Sharp temperature boost\\
 E1            & Extended temperature boost\\
 N1            & No temperature boost; control model\\
 S2            & Similar to S1, but with a rapid change in $n_{\rm e}$\\ 
 S3            & Identical to S1, but with a stricter timestep limit \\ 
 
 \hline
\label{tab:sims}
\end{tabular}
\end{table}

The corresponding volume weighted \HeII and \HI fractions in the
simulations are displayed in the central panel of
Fig.~\ref{fig:ionisation}.  In the S1, E1 and N1 simulations, both \HI
and \HeII reionisation commence at $z=6$.  This choice is deliberate;
these three models are designed to exclude the $8$ per cent increase
in the free electron fraction following \HeII reionisation (see
discussion in \S2).  Instead, they will be used to explore the impact
of differences in the IGM thermal state {\it alone} on the \teff
evolution.  Sudden \HeII reionisation at $z=3.4$ is instead included
in a fourth model, S2 (long-dashed curve), which is tailored to have a
similar temperature at mean density to the S1 model.  The S2
simulation is therefore the most similar to the model used by T02, who
also assumed \HI reionisation at $z=6$ and \HeII reionisation at
$z=3.4$.

The right hand panel in Fig.~\ref{fig:ionisation} displays the
evolution of the slope of the temperature-density relation,
$T=T_{0}\Delta^{\gamma-1}$ (\citealt{HuiGnedin97,Valageas02}), in each
of the simulations.  The power-law index $\gamma$ gradually increases
towards lower redshifts in the E1 and S1 models, asymptotically
approaching the upper limit of $\gamma\sim 1.6$ achieved by the
balance between photo-heating and adiabatic cooling in the low density
IGM.  Note however, the S1 and S2 models, although having very similar
$T_{0}$ values, exhibit different behaviour for $\gamma$.  This is due
to the different ionisation histories adopted in the two models.
During the reionisation of \HeII at $z=3.4$ in the S2 model,
photo-ionisation equilibrium is no longer a good assumption and the
\HeII photo-heating rate is {\it independent} of density
(\citealt{Bolton09}).  This is because the \HeII fraction -- $f_{\rm
  HeII}=n_{\rm HeII}/n_{\rm He} \sim 1$ immediately prior to
reionisation -- is independent of density.  This flattens the
temperature-density relation and lowers the value of $\gamma$.
However, in the S1 model, where the \HeII is already in ionisation
equilibrium with the UV background, the \HeII photo-heating rate is
instead {\it proportional} to the IGM density (\citealt{Theuns05}).
The \HeII fraction is proportional to density (due to higher
recombination rates in denser regions) and the extra energy injected
at $z=3.4$ instead increases $\gamma$.  As we shall see, these
differences will also play a small role (relative to $T_{0}$) in the
simulated \teff evolution.  These values are compared to measurements
of $\gamma$ from \cite{Schaye00}.  Although the error bars are large,
somewhat lower values of $\gamma$ are preferred at $z<3.5$.

\subsection{Time integration in GADGET-2 and Hydra}

We run the fifth and final model, S3, to quantify the effect of the
GADGET-2 time integration scheme on our results.  The study of T02
used hydrodynamical simulations performed with a modified version of
the $\rm P^{3}M-SPH$ code Hydra (\citealt{Couchman95}).  Hydra employs
a single step time integration scheme where all particle positions are
advanced on the minimum timestep required throughout the simulation
volume.  In contrast, GADGET-2 uses individual timesteps for
each particle.  This means that particles in high density regions,
where dynamical timescales are short, have timesteps which are orders
of magnitude smaller than particle timesteps in the lowest density
regions (\citealt{Springel05}).  This enables efficient use of
computational resources and substantially improves code performance.
However, if a sudden boost to the IGM temperature occurs in the middle
of a particle timestep, the energy injected will be smeared over the
timescale corresponding the particle timestep.  In regions where the
dynamical timescale is long, such as the low density IGM which
dominates the transmission in the \Lya forest at $z=3$, this can delay
the impact photo-heating on the gas and could potentially affect our
numerical results.

Unfortunately, forcing a single minimum timestep in GADGET-2 to test
this possibility would be prohibitively expensive.  We instead impose
a stricter upper limit on the maximum timestep size for {\it all}
particles in the S3 model.  In our first four simulations, the maximum
allowed timestep for all particles is $\Delta \ln [1/(1+z)]=0.05$,
although in practice this value will vary and can be up to a factor of
two smaller at any given redshift.  This is equivalent to specifying
the maximum allowed timestep as a fraction of the current Hubble time
in the simulations.  In the S3 model we instead impose a maximum
timestep of $\Delta \ln [1/(1+z)]=0.001$ at $z<4.2$.  This choice
should provide a useful test of the impact of timestepping on the gas
hydrodynamics during a rapid change in the IGM temperature
(V. Springel, private communication).  The thermal and ionisation
history of the S3 model corresponds to the short-dashed curves in
Fig.~\ref{fig:ionisation}.  Note the differences between the thermal
and ionisation histories of the S1 and S3 models are minimal, and that
the curves are almost indistinguishable in all three panels.


\section{The \Lya effective optical depth from hydrodynamical simulations}
\subsection{Construction of synthetic \Lya spectra}

We now turn to analysing the \Lya forest effective optical depth in each
of our simulations.  Synthetic \Lya forest spectra are constructed from
each simulation by extracting randomly selected sight-lines parallel
to the simulation box boundaries at redshift intervals of $\Delta
z=0.05$ over the range $2 \leq z \leq 4$.  At each redshift a total of
$1024$ sight-lines are extracted, each with $1024$ pixels.  Every
pixel in each sight-line has a neutral hydrogen number density
$n_{\rm HI}$, temperature $T$, peculiar velocity $v_{\rm pec}$ and
Hubble velocity $v_{\rm H}$ associated with it.   A standard SPH
interpolation procedure ({\it e.g.}  \citealt{Theuns98}) is used to
extract the first three of these quantities from the simulation data.
In each line of sight with $N$ pixels, the \Lya optical depth in pixel
$i$ is then given by

\begin{equation} \tau(i) = \frac{c \sigma_{\alpha}\delta R }{\pi^{1/2}}  
  \sum_{j=1}^{N} \frac{n_{\rm HI}(j)}{b_{\rm HI}(j)} H(a,x). \label{eq:voigt}
  \end{equation}

\noindent
Here $b_{\rm HI}=(2k_{\rm B}T/m_{\rm H})^{1/2}$ is the Doppler
parameter, $\sigma_{\alpha}=4.48\times 10^{-18}\rm~cm^{2}$ is the \Lya
cross-section, $\delta R$ is the pixel width and  $H(a,x)$ is the
Voigt-Hjerting function (\citealt{Hjerting38})

\begin{equation} H(a,x) = \frac{a}{\pi} \int^{\infty}_{-\infty}
  \frac{e^{-y^{2}}}{a^{2} + (x-y)^{2}} ~dy, \end{equation}

\noindent
where $x = [v_{\rm H}(i) - u(j)]/b_{\rm HI}(j)$, $u(j) = v_{\rm H}(j)
+ v_{\rm pec}(j)$, $a = \Lambda_{\alpha} \lambda_{\alpha} /4\pi b_{\rm
  HI}(j)$, $\Lambda_{\alpha} = 6.265 \times 10^{8} \rm~s^{-1}$ is the
damping constant and $\lambda_{\alpha}=1215.67 \rm~\AA$ is the
wavelength of the \Lya transition.  We use the analytic approximation
for $H(a,x)$ provided by \cite{TepperGarcia06}.

\subsection{Comparison to the \teff evolution observed by FG08b}

\begin{figure}
\begin{center}
\includegraphics[width=0.45\textwidth]{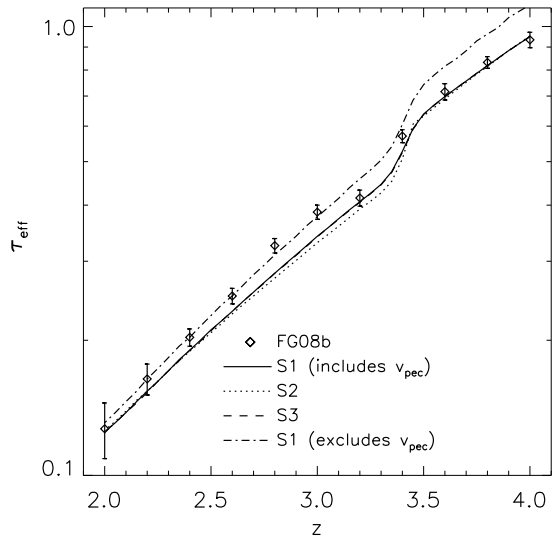}
\vspace{-0.4cm}        
\caption{Comparison of \teff measured from our synthetic \Lya forest
  spectra to the observational data of FG08b (open diamonds).  The
  error bars correspond to the statistical uncertainties only.  The
  synthetic data correspond to the models with a sharp temperature
  boost at $z=3.4$ (S1, solid curve), an $8$ per cent increase in
  the free electron fraction in addition to the sharp temperature
  boost  (S2, dotted curve) and a stricter upper limit on the particle
  timesteps (S3, dashed curve).  The latter model is
  indistinguishable from the S1 data.  A fourth curve (dot-dashed)
  corresponds to the effective optical depth measured from the S1
  model ignoring the effect of peculiar velocities on the \Lya forest.
  As noted by T02, neglecting peculiar velocities raises the absolute
  value of \teff by deblending absorption lines.}
\label{fig:taueff}
\end{center}
\end{figure}

We shall compare our simulation data to the recent direct measurement
presented by \cite{Faucher08} (hereafter FG08b), who found a feature in
the \Lya forest opacity evolution consistent with the one detected
statistically by \cite{Bernardi03}.   FG08b measured \teff over the
redshift range $2\leq z \leq 4.2$ using a sample of 86
high-resolution, high signal-to-noise quasar spectra obtained using
three different instruments; the ESI and HIRES spectrographs on Keck
and the MIKE spectrograph on Magellan. 

We firstly renormalise the optical depths of the synthetic spectra by
the same constant at every redshift, $A=0.87$, to approximately match the
normalisation of the \Lya effective optical depth measured by FG08b.
For a set of spectra with N pixels

\begin{equation}
e^{-\tau_{\rm eff}}= \langle F \rangle =
\frac{1}{N}\sum_{i=1}^{N}e^{-A\tau_{\rm i}}, 
\label{eq:normalise}
\end{equation}

\noindent
where $\tau_{\rm i}$ is the optical depth in each pixel of the
synthetic spectra and $\langle F \rangle$ is the mean observed flux.
From Eq.~(\ref{eq:FGPA}), this renormalisation corresponds to a
straightforward rescaling of $\Gamma_{-12}$ upwards by around $15$ per
cent if all other parameters in the simulation remain fixed.  We have
verified this renormalisation has no impact on the shape of the \teff
evolution.

\begin{figure*}
\centering 
\begin{minipage}{180mm} 
\begin{center}
\includegraphics[width=0.85\textwidth]{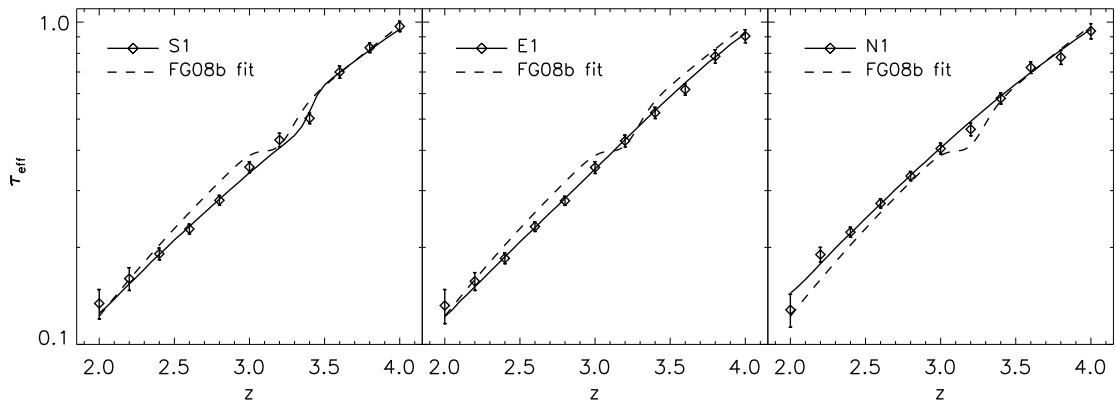}  
\vspace{-0.2cm}
	\caption{A comparison of the \teff evolution measured from
          synthetic \Lya spectra drawn from three of our
          hydrodynamical simulations to the best fitting curve from
          FG08b (dashed curves).   The solid curves in each panel
          display \teff measured from the unprocessed synthetic \Lya
          spectra, while the open diamonds with error bars correspond
          to \teff measured in bins of width $\Delta z = 0.2$ after
          the spectra are processed to closely resemble the FG08b data
          set. The error bars correspond to the standard error of the
          mean.  {\it Left:} The S1 model. {\it Centre:} The E1
          model. {\it Right:} The N1 model.  }
 \label{fig:taueff_real}
\end{center} 
\end{minipage}
\end{figure*}

Fig.~\ref{fig:taueff} shows the comparison between the \teff evolution
measured from our synthetic spectra and the observational data of
FG08b.  Note we have yet not processed the synthetic spectra to
resemble the FG08b data (aside from the overall renormalisation); the
synthetic spectra are noiseless and have a substantially longer path
length than the observational data, but  nevertheless provide a useful
illustration of the \teff evolution in the simulations.  The solid,
dotted and dashed curves display the S1, S2 and S3 models,
respectively.  These data all display the same generic features
observed in the semi-analytic model; a downturn in \teff due to the
reheating of the IGM, followed by a slow recovery driven mainly by
adiabatic cooling.  In all instances it is clear the simulated data do
not exhibit the narrow feature observed in the FG08b data.  

Firstly, note the S3 and S1 models are again almost indistinguishable,
giving us confidence that the adaptive time integration scheme used in
GADGET-2 will not significantly alter our results.  Secondly, a small
difference between the \teff evolution observed in the S1 and S2
models is apparent at $z<3.4$ -- recall that the S2 simulation
includes the $8$ per cent increase in the free electron abundance due
to sudden \HeII reionisation.  As noted in \S2, the extra electrons
will reduce the \HII recombination timescale and thus increase the
\Lya opacity. In a highly ionised IGM the equilibration timescale is
$t_{\rm eq} \simeq f_{\rm HI}t_{\rm rec}$, where $f_{\rm HI}=n_{\rm
  HI}/n_{\rm H}$ is the IGM \HI fraction and the recombination
timescale is

\begin{equation} t_{\rm rec} \simeq \frac{9.4 \times
    10^{9}\rm~yrs}{(1+\chi_{\rm He})\Delta}\left(\frac{T}{2\times
    10^{4}\rm~K}\right)^{0.7}\left(\frac{1+z}{4}\right)^{-3}. \end{equation} 

\noindent
Therefore, for $f_{\rm HI}\simeq 10^{-5.5}$, the equilibration
timescale is already very short at mean density, $t_{\rm eq}\sim
3\times 10^{4}\rm~yrs$ ({\it cf.} $2.9\times 10^{8}\rm~yrs$ between
$z=3.4$ and $z=3$), and the increase in $f_{\rm HI}$ (and hence
$\tau_{\rm eff}$) following the additional $8$ per cent increase in
the electron number density should be very rapid.  However, we instead
find \teff in the S2 model is 3-4 per cent {\it lower} than the S1
model at $3\leq z \leq 3.4$.  In this instance the effect of the extra
electrons on the \Lya opacity is countered by the associated
flattening of the temperature-density relation following sudden,
homogeneous \HeII reionisation.  Recall that most of the \Lya forest
at $z>3$ is dominated by transmission from underdense regions in the
IGM, which become hotter in the S2 model as the temperature-density
relation flattens.  Consequently, including a sudden change in the
ionisation state of \HeII in addition to a large temperature boost
{\it increases} rather than decreases the magnitude of the dip in
\teff at $z<3.4$.  The extra electrons therefore do not help in
reproducing the \teff data in our model.

The fourth, dot-dashed curve corresponds to spectra extracted from the
S1 simulation, but this time excluding the effect of redshift space
distortions induced by peculiar velocities in the IGM.  As noted by
T02, excluding the peculiar velocities tends to deblend \Lya
absorption lines, increasing $\tau_{\rm eff}$.   It is also apparent
the peculiar velocity field flattens the \teff evolution somewhat; the
gradient of the dot-dashed curve is steeper in comparison to the S1
data including peculiar velocities, especially at $z<3$.  These
differences decrease towards lower redshift, indicating line
blending becomes less widespread as the line number density and \Lya
opacity both fall.  However, in contrast to T02 (see their fig. 2),
we find these redshift space distortions are insufficient
to produce the rapid upturn in \teff by $z=2.9$ seen in the
observational data.  The redshift space distortions induced by
peculiar velocities provide no change in the generic shape of
the \teff evolution with redshift.  We have verified that this also
holds for the S3 model with the stricter timestep limit.  This result
will be considered in more detail in \S\ref{sec:hydro}.

\begin{table}
\centering
\caption{Instrumental resolution (FWHM), pixel sizes and signal-to-noise
  properties adopted for the synthetic \Lya forest spectra.  The
  values are based on those reported by FG08b.}
\begin{tabular}
{c|c|c|c}
  \hline
    Instrument & FWHM & Pixel size & S/N per pixel\\  
  \hline
 HIRES         & $6\rm~km~s^{-1}$ & $2\rm~km~s^{-1}$ & 30\\
 ESI           & $33\rm~km~s^{-1}$ & $11\rm~km~s^{-1}$ & 30\\
 MIKE          & $11\rm~km~s^{-1}$ & $2\rm~km~s^{-1}$ & 20\\
 
 \hline
\label{tab:spectra}
\end{tabular}
\end{table}

However, to make a fair comparison to the FG08b data, our synthetic
spectra must be processed to resemble their observational data set as
closely as possible.  Our renormalised synthetic spectra are therefore
also convolved with a Gaussian instrument profile and resampled onto pixels
of the required size.  Gaussian distributed random noise is then added
to each pixel.  The parameters used for this procedure are summarised
in Table~\ref{tab:spectra}, corresponding to the three instruments
used in the FG08b data set.  A random sub-sample of synthetic HIRES,
ESI and MIKE spectra with a total path length corresponding to the
values displayed in fig. 2 of FG08b are then drawn from our
simulated data set in intervals of $\Delta z =0.2$ between $2\leq z
\leq 4$.  

The results are displayed as the open diamonds in
Fig.~\ref{fig:taueff_real} for the S1 (left panel), E1 (central panel)
and N1 (right panel) models.  Following FG08b, the error bars
(statistical only) correspond to the standard error of the mean,
computed by subdividing the synthetic data into chunks $3$ proper Mpc
in length.  The solid curves in each panel correspond to the
underlying \teff evolution measured from the unprocessed synthetic
\Lya spectra.  Our results are compared to the best fitting curve to
the FG08b data, displayed by the dashed line in each panel.

As expected, the synthetic data in Fig.~\ref{fig:taueff_real} now
exhibit some additional scatter due to the variation in cosmic
structure probed by the random sight-lines, but in all instances are
within $1\sigma$ of the underlying \teff evolution.  The S1 data in
the left-hand panel again exhibit the generic behaviour seen in the
semi-analytical model, and are inconsistent with the narrow feature
observed by FG08b.  For comparison, the central panel displays the data
obtained from the E1 model with an extended period of heating ({\it
  e.g.}  \citealt{McQuinn09}), while the right hand panel displays the
N1 model with no additional heating from \HeII reionisation.  In both
instances \teff smoothly evolves with redshift.  This result is in
agreement with the predictions for \teff from the radiative transfer
simulations of \cite{McQuinn09}; an extended period of reheating
during \HeII reionisation will induce no sharp features in the \teff
evolution due to changes in the IGM temperature alone.  

Our detailed simulations therefore indicate that hydrodynamical
effects in the IGM following rapid reheating (but see
\citealt{Bolton09,McQuinn09}) will not aid in reproducing the sharp
dip in the \Lya opacity observed by FG08b.  Instead, our results
suggest any recovery in \teff following sudden reheating will be
driven primarily by adiabatic cooling in the low density IGM,
consistent with the behaviour predicted in our semi-analytical model.


\begin{figure*}
  \centering 
  \begin{minipage}{180mm} 
    \begin{center}

\psfig{figure=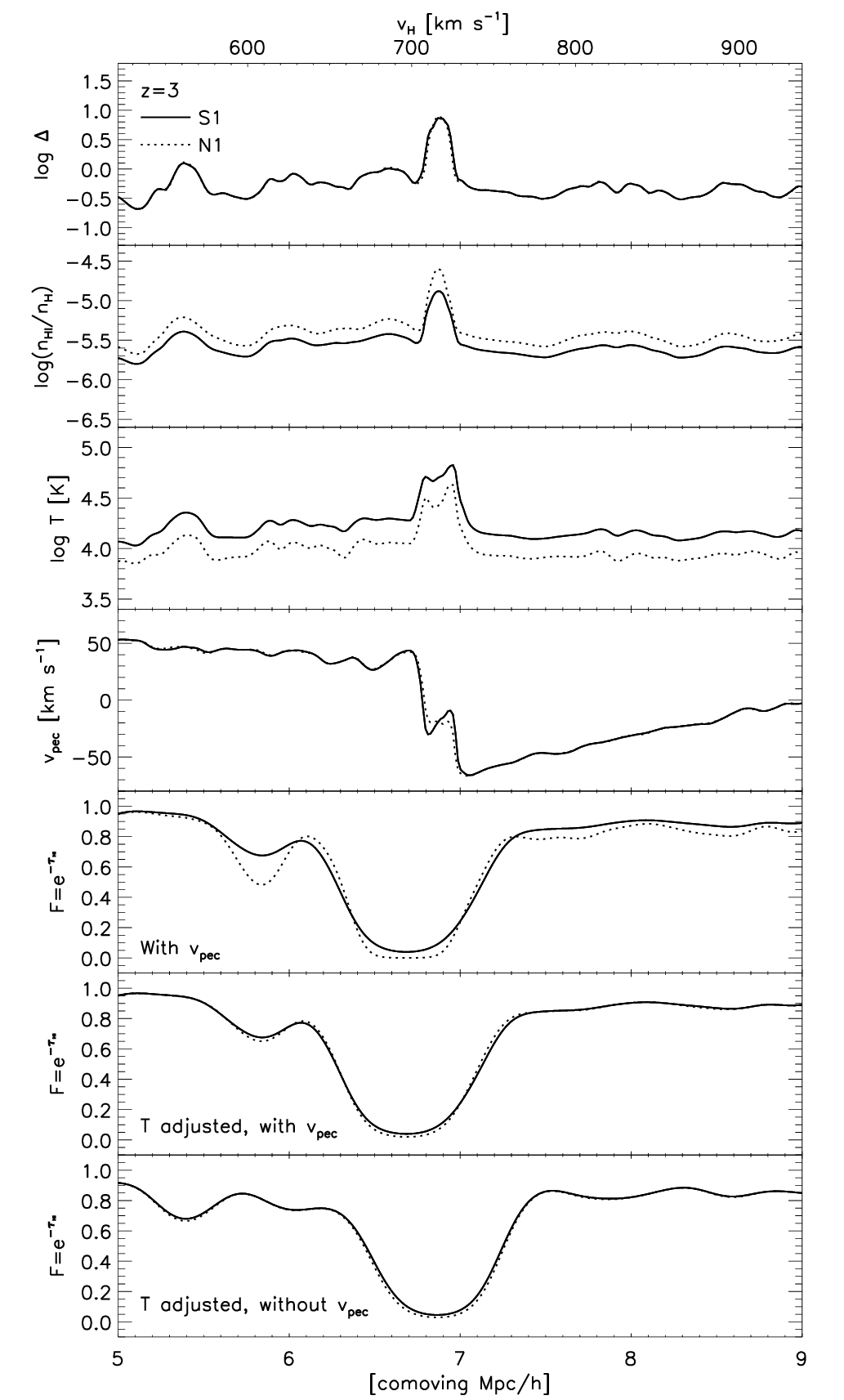,width=0.495\textwidth}
\psfig{figure=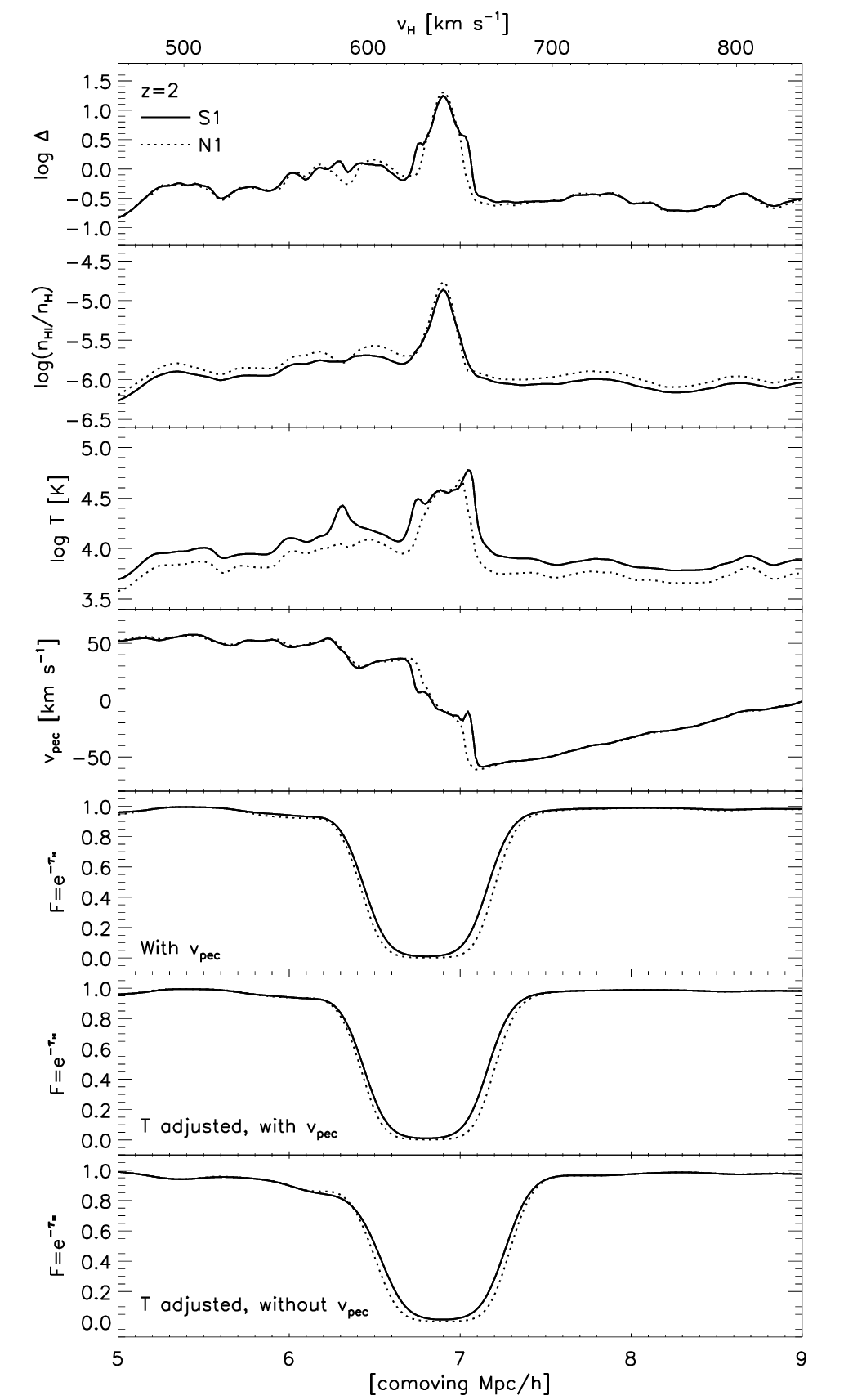,width=0.495\textwidth}

\caption{Comparison between various IGM properties along a subsection
  of a sight-line extracted from identical locations in the S1 (solid
  curves) and N1 simulations (dotted curves) at $z=3$ (left panel) and
  $z=2$ (right panel).  From top to bottom,  the normalised gas
  densities, the \HI fractions, the gas temperatures, the peculiar
  velocities and the resulting \Lya forest spectra, computed in three
  different ways, are displayed.  The spectra in the third row from
  bottom are computed using Eq.~(\ref{eq:voigt}) and include the
  effect of the peculiar velocity field.  In the next row, however,
  the N1 spectrum has been recomputed by fixing the gas temperature in
  the N1 model to be equal to the S1 values and then rescaling the N1
  \HI fraction, such that  $n_{\rm HI}^{\prime} = n_{\rm HI}(T_{\rm
    S1}/T_{\rm N1})^{-0.7}$. This removes most of differences in the
  spectra attributable to Doppler broadening and the \HI fraction.
  The S1 and temperature adjusted N1 spectra both exclude the effect
  of the peculiar velocity field in the bottom row.}
\label{fig:los} 
   \end{center} \end{minipage}
\end{figure*}

\begin{figure*}
  \centering 
  \begin{minipage}{180mm} 
    \begin{center}
            	
      \psfig{figure=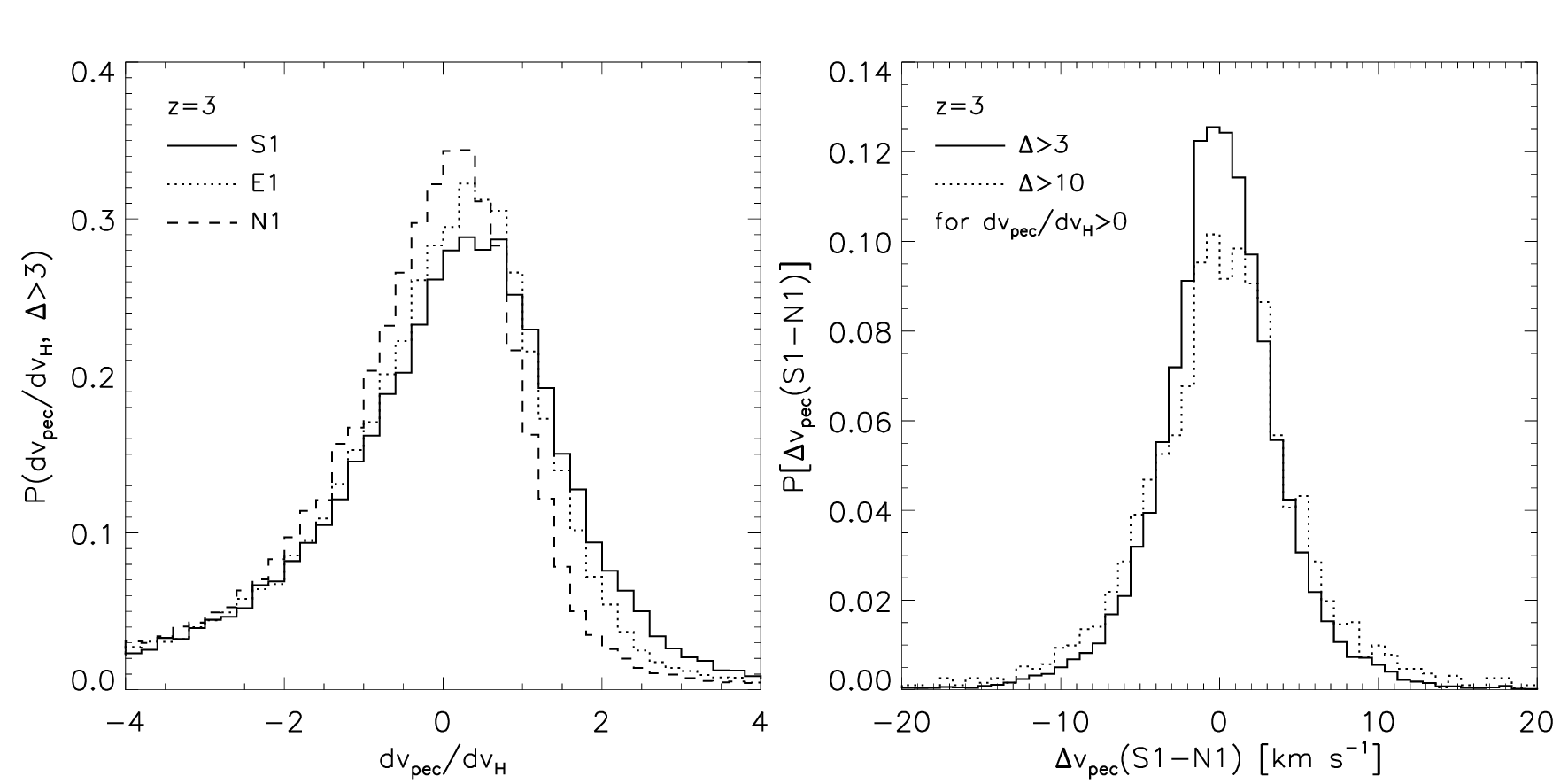,width=0.85\textwidth}  
 
\vspace{-0.2cm}

      \caption{ {\it Left:} Probability distribution of the peculiar
        velocity derivative with respect to the Hubble velocity,
        $v_{\rm H}$, in the S1 (solid curve), E1 (dotted curve) and N1
        (dashed curve) simulations at $z=3$.  The probability
        distribution is for gas with $\Delta>3$ only.  The shift in
        the probability distribution towards more positive gradients
        for progressively hotter models is associated with expansion
        in the higher density regions in the simulations. {\it Right:}
        The probability distribution function of the {\it difference}
        between the peculiar velocity fields in the S1 and N1 models
        in all regions where there is a positive velocity gradient in
        the S1 model ({\it i.e} regions which are expanding).  The
        solid and dotted curves display the distributions for all gas
        with $\Delta>3$ and $\Delta>10$, respectively.  The largest
        values of $\Delta v_{\rm pec}$ correspond to the densest
        regions in the simulations.}
\label{fig:divv}
      \end{center} 
  \end{minipage}
\end{figure*}

\section{Hydrodynamics and redshift space distortions following sudden reheating}
\label{sec:hydro}

In the last section we demonstrated our simulations were unable to
reproduce the narrow feature observed by FG08b.  We now consider this
result in more detail by examining the impact of hydrodynamical
effects and redshift space distortions on the \Lya effective optical
depth evolution.

\subsection{The gas density distribution}

We begin by briefly discussing the effect of sudden reheating on the
IGM density distribution.  Following an injection of energy into the
IGM during reionisation, the subsequent increase in gas pressure will
smooth the IGM density distribution over scales corresponding to the
local Jeans length (\citealt{Schaye01,Pawlik08}).  However, as noted
by \cite{GnedinHui98}, this Jeans smoothing will not occur
instantaneously.  The low density gas responsible for the \Lya forest
absorption will expand on a sound-crossing time, which can be
considerable -- comparable to the Hubble time -- in low density
systems. Indeed, hydrostatic equilibrium is only restored once $t_{\rm
  sc}\sim t_{\rm dyn}$, where $t_{\rm sc}$ and $t_{\rm dyn}$ are the
sound crossing and dynamical timescales respectively
(\citealt{Schaye01}).  The dynamical time $t_{\rm dyn}$ in the low
density IGM is around the Hubble time, as is evident from the
Friedmann equation, $H^{2} = 8 \pi G \rho$.

This can be appreciated in the comparison between sight-line data
drawn from the S1 (solid curves) and N1 (dotted curves) simulations
displayed in Fig.~\ref{fig:los}.  The left panel displays a subsection
of a single sight-line through the simulations at $z=3$, while the
right panel shows the same sight-line at $z=2$. From top to bottom,
the normalised gas densities, the \HI fractions, the gas temperatures,
the peculiar velocities and the resulting \Lya forest spectra,
computed in three different ways, are displayed.  The spectra in the
third row from bottom are computed using Eq.~(\ref{eq:voigt}) and
include the effect of redshift space distortions due to the peculiar
velocity field.  In the next row down, however, the N1 spectrum has
been recomputed by fixing the gas temperature in the N1 model to be
equal to the S1 values and then rescaling the N1 \HI fraction, such
that $n_{\rm HI}^{\prime} = n_{\rm HI}(T_{\rm S1}/T_{\rm N1})^{-0.7}$.
This removes most of differences in the spectra attributable to
Doppler broadening and the \HI fraction, although note the $T^{-0.7}$
scaling for the recombination coefficient is not exact and small
differences will remain.\footnote{In the left hand panel of
  Fig.~\ref{fig:los} prior to this rescaling, the larger difference between
  the S1 and N1 line profiles at $5.8h^{-1}\rm~Mpc$ on comparison to
  the broader lines at $6.7h^{-1}\rm~Mpc$ is because these absorption
  features lie on the linear and logarithmic parts of the curve of
  growth, respectively.}  Finally, in the bottom row the S1 and
temperature adjusted N1 spectra both exclude the effect of the
peculiar velocity field.

It is clear that differences in the \Lya spectra due to the gas
density alone are very small soon after reheating.  The density
distribution at $z=3$ has only been slightly smoothed around the peak
of the overdensity located at $6.9h^{-1}\rm~Mpc$.  In contrast,
pressure smoothing of the gas distribution is clearly apparent by
$z=2$, as there has been sufficient time for the gas distribution to
be altered by the increased pressure in the high density regions.  It
is precisely because of the finite timescale required to achieve
hydrostatic equilibrium that changes in the gas density distribution
due to Jeans smoothing will have a negligible impact on the \teff
evolution by $z=3$ if the IGM is suddenly heated at $z=3.4$.

\subsection{Peculiar velocity gradients}

However, the finite time required to change the gas distribution does
not exclude rapid changes in the peculiar velocity {\it gradients}
responsible for the Jeans smoothing.  Redshift space distortions
associated with these gradients may then induce changes in the \Lya
line profiles (\citealt{Bryan99,Theuns00}).  T02 appealed to these
peculiar velocity gradients to explain the recovery in the \teff
feature observed by \cite{Bernardi03}.  On examination of the peculiar
velocity field in the left panel of Fig.~\ref{fig:los}, the S1 model
does indeed exhibit a positive peculiar velocity gradient in the
centre of the overdensity at $6.9h^{-1}\rm~Mpc$, as it must if the gas
is to expand.  The increased gas pressure has reversed the collapse of
the overdensity which is occurring in the colder N1 model.

This effect of gas temperature on the peculiar velocity gradients is
displayed in more detail in the left hand panel Fig.~\ref{fig:divv},
where the probability distribution of the peculiar velocity derivative
with respect to the Hubble velocity, $\frac{dv_{\rm pec}}{dv_{\rm
    H}}$, in all $1024$ synthetic sight-lines is shown for the S1
(solid curve), E1 (dotted curve) and N1 (dashed curve) simulations.
The distribution is shown for overdense gas with $\Delta>3$ only ({\it
  cf.} fig.~3 in \citealt{Theuns00}).  The shift in the probability
distribution towards more positive gradients for progressively hotter
models is due to expansion in the hotter, high density regions which
have become overpressurised with respect to their cooler surroundings
(\citealt{Bryan99,Theuns00}). The fact that the probability
distributions look qualitatively similar to the data presented in
fig. 3 of \cite{Theuns00} is encouraging, and indicates the higher
temperatures in our GADGET-2 simulations have a similar impact on the
peculiar velocity field.  \cite{Theuns00} used simulations performed
with Hydra in a $2.5h^{-1}$ comoving Mpc box with a gas particle
masses of $1.14 \times 10^{6}\rm~M_{\odot}$ (similar to our mass
resolution).  Note the larger velocity gradients present in our
simulations are most likely due to the additional large scale power
present in our significantly bigger simulation boxes ($15h^{-1}$
comoving Mpc).

In the example sight-line in Fig.~\ref{fig:los}, the maximum
difference between the peculiar velocities in the S1 and N1 models is
$\sim 10\rm~km~s^{-1}$ in the centre of the overdensity.  Although
this difference indeed has some impact on the line profile, the change
in the peculiar velocity field is small on comparison to the line
width, and it is not enough to significantly alter the broad \Lya
line shown in the lower panels.  Once differences attributable to the
different temperatures of the simulations have been scaled out of the
spectra, the S1 and N1 models produce very similar absorption line
profiles.

This anecdotal evidence is displayed more quantitatively in the right
hand panel of Fig.~\ref{fig:divv}, where we plot the probability
distribution of the {\it difference} between the peculiar velocities
in the S1 and N1 simulations for all pixels with positive peculiar
velocity gradients and $\Delta>3$ (solid curve) or $\Delta>10$ (dotted
curve) in the S1 data.  In the majority of these regions there is in
fact only a small change in the peculiar velocity field, with the
largest differences ($\sim 10\rm~km~s^{-1}$) associated with the
rarer, high density peaks which produce broad lines like the one shown
in Fig.~\ref{fig:los}.  Since an absolute change in the peculiar
velocity of a few $\rm km~s^{-1}$ is small in comparison to typical
line widths of $20\rm~km~s^{-1}$ ({\it e.g.}  \citealt{Kim02}), this
explains why the impact on \teff is correspondingly small.   Instead,
it is the {\it instantaneous} temperature of the IGM which primarily
influences the \teff evolution in our simulations.

These findings differ from those of T02, who found changes in the
peculiar velocity gradients were partially responsible for inducing
the recovery in \teff observed by \cite{Bernardi03}.   The discrepancy
between these results may be due to differences between the numerical
methods used.  We have tested the impact of timestepping on
simulations, and we find our results are robust in this respect.
However, we cannot be absolutely certain that other numerical effects
do not play a role, and ultimately we can only speculate on the origin
of this difference.  Ideally, an independent numerical study is
required to verify or refute our claims.  In agreement with T02,
however, we do indeed find that line blending due to the peculiar
velocity field lowers $\tau_{\rm eff}$, and that higher temperatures
steepen the peculiar velocity gradients in overdense regions, giving
us confidence that our simulations are at least broadly consistent
with T02.  We find these effects are nevertheless insufficient to
reproduce the narrow feature observed in the \teff evolution,
suggesting that a sudden increase in the IGM temperature at $z=3.4$
following \HeII reionisation is unable to adequately explain the \teff
data.


\section{On the possible origin of the observed \teff feature}

Thus far we have established that even in the presence of a sudden
increase in the IGM temperature at $z=3.4$ (but see
\citealt{Bolton09,McQuinn09}), the narrow dip observed in the \Lya
effective optical depth by FG08b cannot be reproduced in our
hydrodynamical simulations of the \Lya forest.  However, before
proceeding it is worth briefly emphasising that the purported
``narrowness'' of the $\tau_{\rm eff}$ feature is a somewhat
model-dependent statement. For instance, if we consider the $\tau_{\rm
  eff}-z$ plane, FG08b characterised the width of the feature by
fitting a power-law and a Gaussian ``bump''.  This significance of the
bump then depends on how good an approximation a power law is to the
underlying evolution of the IGM opacity excluding the effects
associated with \HeII reionisation.  In our simulations without sudden
heating, $\tau_{\rm eff}$ does indeed evolve smoothly with redshift,
indicating this should be a reasonable approximation if the redshift
evolution of the ionising background is gradual.  In contrast, the
FG08b feature stands out visually, and more objective measures we have
applied to the data ({\it e.g.}, a regularised derivative of the data
as used in edge-finding algorithms) tend to confirm that impression.
The fact that it is seen in at least three independent data sets
further hints that it is not a data artifact. 

Additionally, a decomposition into the required evolution in the
ionising background (as in the next section) will also depend on the
assumed IGM thermal evolution.  The \Lya opacity scales as $\tau
\propto T^{-0.7}/\Gamma_{\rm HI}$; the effect of gas temperature and
the photoionisation rate on $\tau_{\rm eff}$ are impossible to
disentangle without independent estimates for one or the other.  In
this work we have demonstrated that sudden heating on its own cannot
reproduce the $\tau_{\rm eff}$ feature.  However, a sudden global
heating event followed by a gradual downturn in the photo-ionisation
rate at $z<3.2$ could still reproduce the data.  On the other hand,
there is good reason to suppose that such a sudden, global boost to
the IGM temperature is unlikely in the first place.  Our previous work
has demonstrated that \HeII reionisation likely produces too little
heating to produce a substantial opacity decrease over the short
timescale required \citep{Bolton09}.  In the next section we therefore
assume the temperature boost during \HeII reionisation occurs over an
extended period of time (our E1 model).  This model also resembles the
results from recent radiative transfer simulations performed by
\cite{McQuinn09}.  We therefore dispense with rapid changes in the
temperature entirely, and now turn to discuss the remaining
possibility for the origin of the \teff dip: a narrow peak in the
metagalactic hydrogen photo-ionisation rate.

\subsection{The hydrogen photo-ionisation rate required by the FG08b
  \teff evolution}

Many studies have used simulations of the \Lya forest, combined with
measurements of $\tau_{\rm eff}$, to place constraints on the
metagalactic hydrogen photo-ionisation rate, $\Gamma_{-12}=\Gamma_{\rm
  HI}/10^{-12}\rm~s^{-1}$ ({\it e.g.}
\citealt{Rauch97,McDonaldMiraldaEscude01,CenMcDonald02,Schaye03,MeiksinWhite04,Tytler04,Bolton05,Jena05}).
We use the same procedure in this work to estimate $\Gamma_{-12}$ from
our E1 simulation, which was constructed to have a similar thermal
history at mean density to the recent radiative transfer simulations
of \cite{McQuinn09}.  We use Eq.~(\ref{eq:normalise}) and the best
fit\footnote{Specifically, we use the best fit FG08b present for their
  \teff measurements in redshift bins of width $\Delta z=0.2$ when
  using the \cite{Schaye03} metal correction.} to \teff obtained by
FG08b for this purpose.  Following from Eq.~(\ref{eq:FGPA}), the
photo-ionisation rate which reproduces the \teff fit is given by
$\Gamma_{-12}(z)=\Gamma_{\rm HM}(z)/A(z)$, where $\Gamma_{\rm HM}(z)$
is the photo-ionisation rate from the HM01 UVB model divided by
$10^{-12}\rm~s^{-1}$.

\begin{figure}
\begin{center}
  \includegraphics[width=0.45\textwidth]{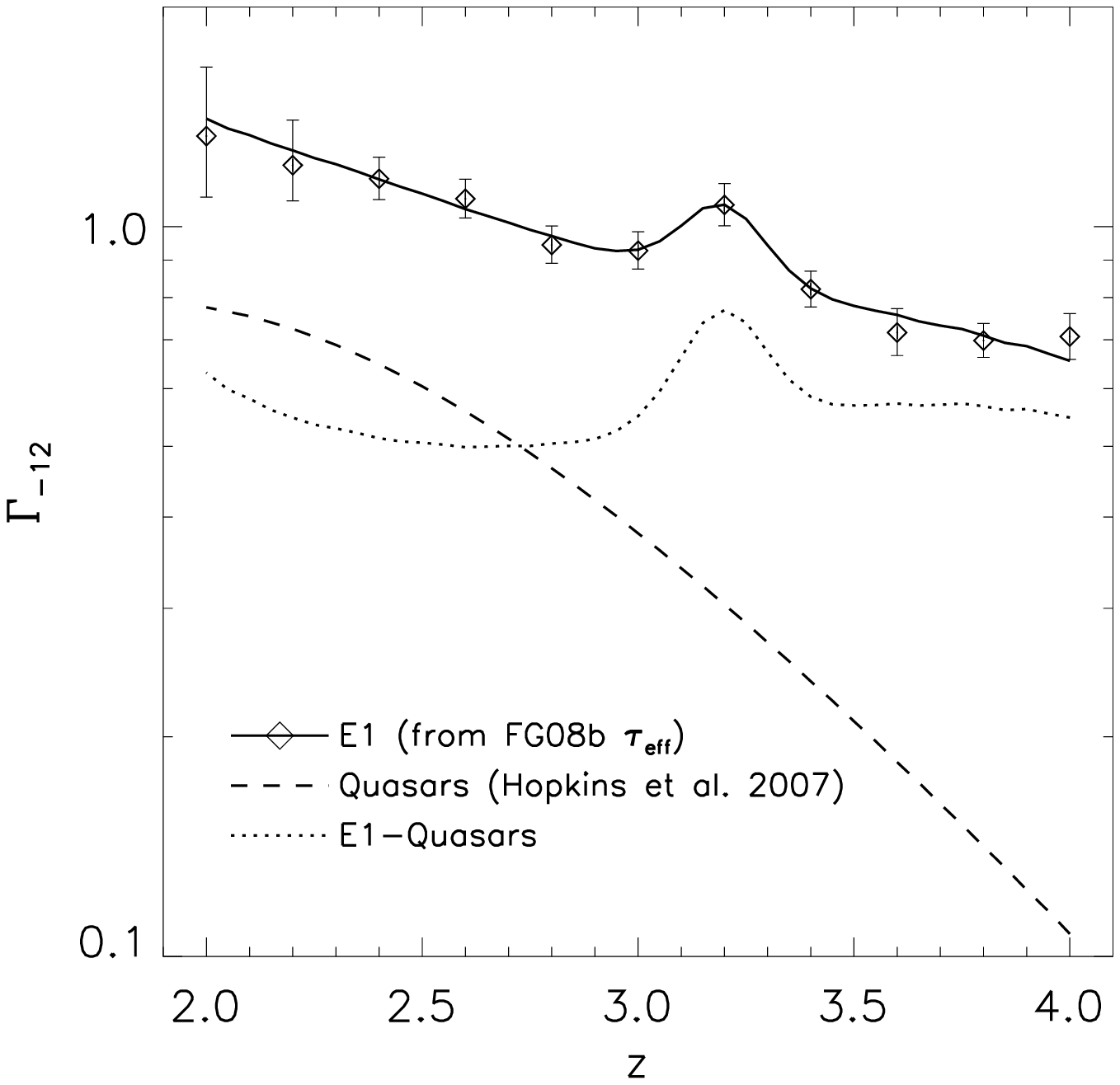}
\vspace{-0.4cm}        
\caption{The solid curve corresponds to the hydrogen photo-ionisation
  rate, $\Gamma_{-12}=\Gamma_{\rm HI}/10^{-12}\rm~s^{-1}$, required to
  reproduce the fit to \teff evolution presented by FG08b.  The narrow
  peak in $\Gamma_{-12}$ at $z=3.2$ corresponds to the dip in the
  FG08b \teff evolution.  The dashed curve displays the expected
  contribution of quasars to the total $\Gamma_{-12}$, based on the
  \citet{Hopkins07} model for the quasar luminosity function, and the
  dotted line corresponds to the difference between the total
  $\Gamma_{-12}$ required by \teff and the quasar only contribution.
  To give an indication of the statistical uncertainty in this peak,
  the open diamonds with error bars are derived from the FG08b \teff
  data points including their $1\sigma$ statistical errors.}  

\label{fig:gamma}
\end{center}
\end{figure}

The result of this procedure is displayed as the solid curve in
Fig.~\ref{fig:gamma}.  A gradual increase in $\Gamma_{-12}$ towards
lower redshift is required, and the prominent bump centred at $z=3.2$
is responsible for the narrow \teff feature.  The open diamonds with
error bars are derived from the FG08b \teff data points including
their $1\sigma$ statistical errors; these give an indication of the
statistical uncertainty in this peak. 

The dashed curve shows the expected contribution from quasars to
$\Gamma_{-12}$. We compute this using the recent \cite{MadauHaardt08}
parameterisation for the \cite{Hopkins07} comoving quasar emissivity
at the \HI Lyman limit, $\nu_{\rm L}$, assuming a power law spectrum
with  $\epsilon_{\nu}^{\rm Q} =\epsilon_{\rm L}(\nu/\nu_{\rm
  L})^{-1.6}$ ({\it e.g.}  \citealt{Telfer02}).  The expected
contribution to the photo-ionisation rate from quasars is then

\begin{equation}  \Gamma_{\rm -12}^{\rm Q} = \frac{(1+z)^{3}}{10^{-12}
   \rm~s^{-1}}\int_{\nu_{\rm L}}^{\infty}\frac{\epsilon_{\nu}^{\rm
      Q}\lambda_{\nu}\sigma_{\nu}}{h_{\rm p}\nu}
  d\nu, \label{eq:Qgamma} \end{equation}

\noindent
where the photo-ionisation cross-section $\sigma_{\nu}=6.3\times
10^{-18}{\rm~cm^{2}}(\nu/\nu_{\rm L})^{-3}$ and the mean free path
$\lambda_{\nu}=\lambda_{\rm L}(\nu/\nu_{\rm L})^{1.5}$; the latter
assumes the \HI column density distribution $f(N_{\rm HI},z)\propto
N_{\rm HI}^{-1.5}$ ({\it e.g.} \citealt{Petitjean93,MiraldaEscude03}).
We take $\lambda_{\rm L}=50[(1+z)/4]^{-4}$ proper Mpc, which is
towards the lower end of the range recently advocated by
\cite{Faucher08b}.   Note, however, that the  expected contribution to
the photo-ionisation rate from quasars depends on the uncertain quasar
luminosity function (particularly the faint end slope), mean free path
and ionising spectrum, as well as radiative transfer effects (for
instance, reprocessed radiation can contribute significantly;
\citealt{HaardtMadau96}).

The dotted curve displays the difference between the total
photo-ionisation rate and the contribution from quasars alone,
$\Gamma_{-12}^{\prime}=\Gamma_{-12}-\Gamma_{-12}^{\rm Q}$.  
This curve remains almost flat for $2\leq z \leq 4$, with
$\Gamma_{-12}^{\prime}\simeq 0.55$, aside from the narrow bump
($\Delta z\simeq 0.4$) which peaks with $\Gamma_{-12}^{\prime}\simeq
0.75$ at $z=3.2$.  Thus, in order to reproduce the observed $\tau_{\rm
  eff}$, we require a boost of around $35-40$ per cent in
$\Gamma_{-12}^{\prime}$ over a short redshift interval. 

Finally, note that although the FG08b statistical errors are
relatively small, the normalisation of $\Gamma_{-12}$ is still
somewhat uncertain.  In particular, although the $\Gamma_{-12}$ values
we derive from our E1 simulation are consistent with other estimates
from hydrodynamical simulations (\citealt{Tytler04,Bolton05,Jena05}),
they are systematically higher by up to a factor of two in comparison
to the recent estimates presented by \cite{Faucher08b} using the same
\teff data.  This is partially because \cite{Faucher08b} do not model
the \Lya forest in detail, instead obtaining analytical estimates for
$\Gamma_{-12}$ using the FGPA combined with the IGM density
distribution derived by \cite{MiraldaEscude00} (see Eqs.~\ref{eq:FGPA}
and \ref{eq:pvd}).  Systematic uncertainties on $\Gamma_{-12}$ due to
other parameters, such as the IGM temperature which is still poorly
constrained, are also large (\citealt{Bolton05}).  Thus, although the
overall normalisation remains somewhat uncertain, the {\it shape} of
these curves should be fairly robust, and the requirement for a peak
in $\Gamma_{-12}$ remains so long as the volume weighted IGM
temperature evolves slowly (\citealt{McQuinn09}).

\subsection{Possible causes of a peak in the hydrogen photo-ionisation rate}
\label{section:ionizing_background}

We have found that, even if the IGM temperature increases suddenly, we
cannot reproduce a narrow feature in the optical depth.\footnote{ The
  exception to this is if the photo-ionisation rate subsequently drops
  at $z<3.2$ following a sudden temperature increase.  However, as
  discussed earlier, recent studies indicate that a rapid global
  temperature boost is unlikely in the first place
  (\citealt{Bolton09,McQuinn09}).}  Thus a narrow peak at $z=3.2$ in
the otherwise approximately constant value of
$\Gamma_{-12}^{\prime}=\Gamma_{-12}-\Gamma_{-12}^{\rm Q}$ from $2\leq
z \leq 4$ is instead required to reproduce the \teff evolution
measured by FG08b.  This could be associated with a sharp modulation
in the ionising emissivity or the mean free path of ionising photons,
or indeed a combination of both effects (see eq.~\ref{eq:Qgamma}).
However, while appealing to the stellar contribution to the
photo-ionisation rate for an increase in the emissivity is in
principle acceptable given the uncertainties involved in deriving the
stellar ionising rate from observations at $z>3$ ({\it e.g.}
\citealt{Madau99,MiraldaEscude03,BoltonHaehnelt07b}), there is no
apparent reason why such a narrow peak should occur at $z=3.2$.

Instead, we suggest it is more likely that any modulation of the
hydrogen photo-ionisation rate is associated with dense \HeII and \HI
Lyman limit systems.  Unlike the low density IGM, such systems have
dynamical, cooling and recombination times which are comparable or
shorter than the $\sim 10^{8}$ years associated with the width of the
$\tau_{\rm eff}$ feature.  They can modulate the hydrogen ionising
background, either by changing the hydrogen ionising emissivity (by
the reprocessing of \HeII ionising photons into \HeII Lyman series,
Balmer, or two-photon emission, all of which can ionise hydrogen), or
the mean free path of ionising photons, since their opacity will be
altered by changes in their size or temperature\footnote{The heating
  and expansion of Lyman limit systems was also briefly discussed by
  \cite{McQuinn09}.}.  Furthermore, a sharp change in the emissivity
or opacity will then be imprinted on the ionising background on a
timescale comparable to the mean free time of an ionising photon, or
$\Delta z \sim (dN/dz)^{-1} [\Gamma(\beta-1)]^{-1} \sim 0.24$ where
$dN/dz=3.3 [(1+z)/5]^{1.5}$ is the abundance of \HI Lyman limit
systems \citep{StorrieLombardi94}, and we assume $dN/dN_{\rm HI}
\propto N_{\rm HI}^{-\beta}$ with $\beta=1.5$.  This is certainly well
within the range required to explain the feature.  We now proceed to
give some brief examples of these effects.

\begin{itemize}

\item{\it Helium recombination radiation}

A large fraction of the hydrogen ionising background may be radiation
from higher frequencies that is reprocessed by dense systems. For
instance, \cite{HaardtMadau96} find that $\sim 40$ per cent of
$\Gamma_{\rm HI}$ comes from reprocessed radiation at $z \sim 3$;
similarly \citep{Fardal98} find that about $\sim 20$ per cent of
$\Gamma_{\rm HI}$ comes from reprocessed radiation.  This has two
components: recombinations of hydrogen to the ground state, and
reprocessing of \HeII ionising photons into \HeII Lyman series,
Balmer, or two-photon emission, all of which can ionise hydrogen. The
latter obviously undergoes rapid evolution during the process of \HeII
reionisation, particularly toward the tail end of reionisation when
the mean ionising background can rise rapidly (although there could be
large fluctuations in the background throughout the reionisation
process; \citealt{Furlanetto08b}).  On the other hand, recent
calculations indicate the contribution of recombination radiation to
the UV background could be smaller than previous estimates
(\citealt{Faucher09}).

Another potentially important effect comes from the reprocessing of
\HeII Lyman series photons in an optically thick IGM
\citep{MadauHaardt08}. Lyman series photons between Ly$\beta$ (at
$E=3.56$~Ry) and the Lyman limit (4 Ry) are degraded to \HeII
Ly$\alpha$ photons, Balmer or lower frequency radiation.  The
magnitude of this effect depends strongly on the opacity of the IGM,
which of course evolves rapidly during \HeII reionisation.  While this
may not appear to be particularly significant (since such photons can
already ionise hydrogen), the degradation to lower energies implies
that the hydrogen photo-ionisation rate can be significantly
modulated.  In particular, selection rules forbid Ly$\beta$ photons
from being converted into Ly$\alpha$ photons, and the reduction of
two-photon emission from Ly$\beta$ reprocessing at the end of \HeII
reionisation could reduce the hydrogen ionisation rate, helping to
produce the downturn in the photo-ionisation rate. 
\\

\item{\it The size of Lyman limit systems}

In this work, we have thus far argued that hydrodynamic effects
associated with the heating of the IGM during \HeII reionisation have
little effect on $\tau_{\rm eff}$.  While this is true of the
low-density gas responsible for most of the absorption in the \Lya
forest, it is not true of higher density systems which have smaller
dynamical timescales, $t_{\rm dyn}\propto \Delta^{-1/2}$.  Since  \HI
Lyman limit systems (LLSs) dominate the opacity of the
post-reionisation IGM, a sudden change in the sizes of these systems
can then modulate the mean free path of \HI ionising photons.

We may consider this as follows.  At the end of \HeII reionisation,
although most of the IGM has been reheated, self-shielded \HeII
LLSs, in which the helium is still predominantly in
the form of He~$\rm \scriptstyle II$, will still remain.
Self-shielding to \HeII ionising photons occurs for \HeII column
densities $N_{\rm HeII}>6.7 \times 10^{17}\rm~cm^{-2}$.  Assuming the
size of an absorber with overdensity $\Delta$ is the local Jeans
length (\citealt{Schaye01}), this column density corresponds to a
characteristic overdensity 

\begin{equation} \Delta \simeq 56
  \left(\frac{T}{10^{4}\rm~K}\right)^{2/15}\left(\frac{1+z}{4}\right)^{-3}\Gamma_{-14}^{2/3}, \end{equation}

\noindent
where $\Gamma_{-14}=\Gamma_{\rm HeII}/10^{-14}{\rm s^{-1}}$ is the
\HeII photo-ionisation rate.  The \HI column density of an absorber
with overdensity $\Delta$ is (\citealt{Schaye01})

\begin{equation} N_{\rm HI}\simeq 3.8 \times 10^{16} {\rm cm^{-2}}
  \frac{\Delta^{3/2}}{\Gamma_{-12}}
  \left(\frac{T}{10^{4}\rm~K}\right)^{-1/5}\left(\frac{1+z}{4}\right)^{9/2}. \end{equation}

\noindent
Self-shielding to \HeII ionising photons therefore happens at an \HI
column density of $N_{\rm HI} \sim 1.6 \times 10^{16} {\rm~cm^{-2}}
(\Gamma_{-14}/\Gamma_{-12})$, well short of the column density of
$N_{\rm HI} \sim 1.6 \times 10^{17} {\rm cm^{-2}}$ associated with \HI
LLSs. Thus, at face value it may seem that \HeII reionisation cannot
substantially affect \HI LLSs, and hence the  mean free path of \HI
ionising photons. However, \HeII LLSs are also likely to be associated
with the less dense outer regions of \HI LLSs. A strong increase in
external pressure due to the reheating of the IGM during \HeII
reionisation then means that the system will no longer be in
hydrostatic equilibrium.  The pressure gradient compresses the
gas,\footnote{Note that in our optically thin hydrodynamical
  simulations the self-shielding of dense systems is not modelled.
  This effect is therefore absent in our simulation data.} until
pressure equilibrium is once again restored (a similar mechanism has
been invoked for globular cluster formation, {\it e.g.}
\citealt{cen01_gc}). As a result, \HI LLSs could decrease in size, and
increase in density. Alternatively, if hard photons can penetrate and
heat the gas in the high density regions ({\it e.g.}
\citealt{Bolton09}), the \HI LLS could instead expand. 

For simplicity, let us for the moment assume that any change in
density occurs isothermally (see below for more discussion). If the
absorber expands or contracts to a new overdensity $\Delta \rightarrow
f \Delta$, then (since $n_{\rm HI} \propto \Delta^{2}$ and $R \propto
\Delta^{-1/3}$), $N_{\rm HI} \propto f^{5/3}$, while the cross-section
of the absorber $\sigma \propto f^{-2/3}$. Assuming that $f$ is
independent of $\Delta$ (the opacity is dominated by \HI LLSs, so the
result will be most affected by $f$ for such systems), the mean free
path of ionising photons will be altered by a factor

\begin{equation}
\frac{\lambda}{\lambda_{\rm L}} = \frac{\int_{0.1}^{\infty} \tau^{-1.5} \left[ 1 - {\rm exp(-\tau)} \, d\tau \right]}{\int_{0.1}^{\infty} f^{-2/3} \tau^{-1.5} \left[ 1 - {\rm exp}(- f^{5/3} \tau) \, d\tau \right]},
\end{equation}
where the lower limit $\tau\sim 0.1$ is the \HI optical depth at which
systems self-shield from \HeII ionising radiation.  This is a
non-monotonic function with a minimum at $f\sim 0.3$; however, it does
not exceed unity until $f>1$. Thus, if systems compress, the hydrogen
photo-ionisation rate falls; if they expand, it increases. However,
more quantitative exploration of this possibility requires careful
simulation of the LLS with coupled hydrodynamics and radiative
transfer, particularly since the equilibrium temperature also strongly
affects opacity.  In addition, a caveat to this argument is that
although the dynamical timescale of an individual Lyman limit system
is comparable to that required for the modulation of $\Gamma_{\rm
  HI}$, a globally averaged change in the mean free path still
requires the size of all the Lyman limit systems to change over a
short time interval.\\

\item{\it The temperature of Lyman limit systems} 

Yet another possibility is that the temperature of \HI LLSs itself
changes during \HeII reionisation, altering their opacity. A rapid
increase in the injection of photons at 3 Ry from reprocessed
radiation implies that the ionising background hardens significantly,
which could result in a significant change in the equilibrium
temperature. This could be further modulated by the hydrodynamic
effects mentioned above.  Note that the equilibrium temperature of
LLSs is a non-trivial function of density in photo-ionisation
equilibrium, particularly in the presence of metal line cooling
\citep{Wiersma08}. Generally, as a system becomes denser, the
equilibrium temperature falls, since the efficiency of cooling
increases as the ionisation parameter falls
\citep{Efstathiou92,Wiersma08}. This exacerbates the increase in
opacity due to the density increase. Non-monotonic evolution of the
temperature of the LLSs (which have short cooling times since they
cool radiatively, rather than adiabatically) could thus modulate their
opacity, as well as the mean free path and the \HI photo-ionisation
rate.\\

\end{itemize}


\section{Conclusions}

We have used a semi-analytic model of inhomogeneous \HeII reionisation
and high resolution hydrodynamical simulations of the \Lya forest to
investigate the impact of sudden reheating on the evolution of the
\Lya forest effective optical depth.  Our semi-analytic model
indicates that any injection of energy into the IGM during
inhomogeneous \HeII reionisation will produce a well understood and
generic evolution in $\tau_{\rm eff}$, where a
reduction in the opacity from $z=4$ to $z=3$ is followed by a gradual,
monotonic recovery driven largely by adiabatic cooling in the low
density IGM.  This behaviour is inconsistent with the narrow dip
($\Delta z=0.4$) of around 10 per cent in \teff at $z=3.2$ which has
now been detected by three independent observational studies
(\citealt{Bernardi03,Dallaglio08,Faucher08}). 

However, our semi-analytic model does not include a detailed
reconstruction of the \Lya forest.  We therefore also analyse five
high resolution hydrodynamical simulations of the IGM to investigate
the effect of various thermal histories on the \Lya effective optical
depth.  We find that sudden reheating at $z=3.4$ results in a sharp
decrease in $\tau_{\rm eff}$, although we note that such a large,
sudden increase in the IGM temperature is nevertheless unlikely to
occur over the entire IGM at once (\citealt{Bolton09,McQuinn09}).
This assumption may nevertheless be appropriate in localised regions
around quasars with hard spectra, and in small volumes similar to the
box size of our hydrodynamical simulations. 

Although the assumption of sudden, homogeneous reheating does indeed
successfully reproduce the initiation of the \teff dip observed by
FG08b, our simulations are still unable to account for the rapid {\it
  recovery} of the narrow dip in the \teff evolution by $z=2.9$.  The
effect of the 8 per cent increase in the free extra fraction following
\HeII reionisation on \teff is small, and any resulting increase in
\teff is instead countered by the simultaneous flattening of the
power-law temperature-density relation during instantaneous,
homogeneous \HeII reionisation at $z=3.4$.  Redshift space distortions
in the \Lya forest attributable to the response of the gas to the
extra energy injected into the IGM are also unable to account for the
observed recovery of $\tau_{\rm eff}$.  We find that sudden reheating
does indeed produce larger, more positive peculiar velocity gradients
corresponding to regions of newly expanding gas, but the absolute
changes in the velocity field are generally small in comparison to
typical line widths, except in the most overdense regions in the
simulation.  Such regions are, however, rare and thus contribute
little to the average opacity.  We have tested the robustness of this
result with respect to the time integration scheme employed in
GADGET-2, and we find this has little impact on our numerical results.
In contrast, if \HeII reionisation is an extended process, then in
agreement with the recent study by \cite{McQuinn09} we find the \Lya
effective optical depth will evolve smoothly with redshift.

As a consequence, we must instead appeal to a narrow peak in the
metagalactic hydrogen photo-ionisation rate at $z=3.2$ to reproduce
the \teff feature in our simulations. This could potentially be
modulated by Lyman limit systems, which have recombination, cooling
and dynamical times comparable to or less than the timescale
associated with the width of the feature. In particular, we suggest
that radiative transfer effects from \HeII reionisation itself could
be responsible, either by altering the emissivity of reprocessed \HeII
recombination photons, or by changing the opacity of \HI Lyman limit
systems and hence the mean free path of ionising photons.  However,
further detailed investigation is still required to establish the
origin of this intriguing feature in the redshift evolution of the
\Lya forest opacity.

\section*{Acknowledgements}

We thank George Becker, Claude-Andr{\'e} Faucher-Gigu{\`e}re, Martin Haehnelt,
Joop Schaye, Tom Theuns and Matteo Viel for helpful discussions during
the course of this work.  We are also very grateful to Volker Springel
for his advice and for providing GADGET-2.  The hydrodynamical
simulations used in this work were performed using the SGI Altix 4700
supercomputer COSMOS at the Department of Applied Mathematics and
Theoretical Physics in Cambridge.  COSMOS is a UK-CCC facility which
is sponsored by SGI, Intel, HEFCE and STFC.  This research was also
supported in part by the National Science Foundation under Grant
Nos. PHY05-51164 (JSB, through the MPA/KITP postdoctoral exchange
programme), and AST-0829737 (SRF), the David and Lucile Packard
Foundation (SRF), and NASA grant NNG06GH95G (SPO).  JSB thanks the
staff at the Kavli Institute for Theoretical Physics, Santa Barbara,
for their hospitality during the early stages of this work.

\bibliographystyle{apj}

\end{document}